\documentclass[aps,prb,twocolumn,superscriptaddress,floatfix,10pt]{revtex4-2}
\usepackage{verbatim}
\usepackage{lipsum}
\usepackage{physics}
\usepackage{bbold}
\usepackage[percent]{overpic}
\usepackage{xcolor}
\usepackage{orcidlink}
\usepackage{graphicx}
\usepackage{amsmath,amssymb,amsfonts,nccmath}
\usepackage{bm}
\usepackage{mathtools}
\usepackage{subcaption}
\usepackage{caption}
\usepackage{empheq}
\newcommand{\panellab}[1]{\put(3,66){\colorbox{white}{\bfseries #1}}}
\usepackage{graphicx,graphics}

\usepackage{hyperref}
\hypersetup{
colorlinks=true,
linkcolor=blue,
filecolor=blue,
citecolor = blue,      
urlcolor=blue,
}

\usepackage{ragged2e}
\makeatletter
\long\def\@makecaption#1#2{%
  \vskip\abovecaptionskip
  \sbox\@tempboxa{#1. #2}%
  \ifdim\wd\@tempboxa>\hsize
    {\justifying\normalfont\noindent #1. #2\par}
  \else
    \global\@minipagefalse
    \hb@xt@\hsize{\hfil\box\@tempboxa\hfil}
  \fi
  \vskip\belowcaptionskip}
\makeatother

\begin{document}

\preprint{APS/123-QED}

\title{Angular Momentum Transfer in Magnetic Weyl Semimetal Spheres}

\author{Adriano Biondo \orcidlink{0009-0006-1263-9233}}
  \affiliation{Dipartimento di Fisica e Astronomia ``Ettore Majorana'', Universit\`a di Catania, Via S. Sofia 64, I-95123 Catania,~Italy.}
  %
\affiliation{Scuola Superiore di Catania, Via Valdisavoia 9, I-95123 Catania, Italy.}

\author{Francesco M. D. Pellegrino \orcidlink{0000-0001-5425-1292}}%
\affiliation{Dipartimento di Fisica e Astronomia ``Ettore Majorana'', Universit\`a di Catania, Via S. Sofia 64, I-95123 Catania,~Italy.}
\affiliation{INFN, Sez.~Catania, I-95123 Catania,~Italy.}
%

\begin{abstract}
In this work, we investigate the problem of electromagnetic angular momentum within the framework of topological materials, described by axion electrodynamics.
 Specifically, we calculate the electromagnetic angular momentum for a spherical sample of magnetic Weyl semimetal in the presence of a point charge. We then analyze how, by moving the point charge quasi-statically, the angular momentum stored in the electromagnetic field can be converted into mechanical angular momentum of the sphere. Finally,  we derive analytical expressions for the resulting angular velocity and angular displacement, demonstrating that the effect is enhanced in the Weyl semimetal compared with a topological insulator.
\end{abstract}
\maketitle

\section{Introduction}\label{sec:intro}

The fact that the electromagnetic field carries not only energy and linear momentum, but also angular momentum, is a direct consequence of classical electrodynamics \cite{jackson_classical_electrodynamics}. Although this property is well established, its physical implications remain particularly intriguing, especially in apparently stationary systems.
In such situations, even if the matter is initially at rest, mechanical motion may arise, provided that the angular momentum is initially stored in the electromagnetic field and subsequently transferred to the material bodies.
    \begin{figure}[t]
     \centering
     \includegraphics[width=0.60\columnwidth]{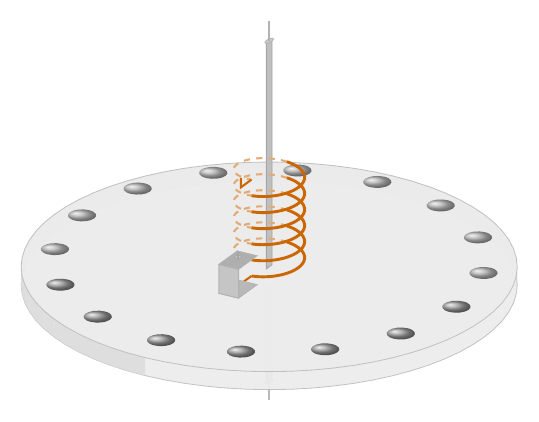}
     \caption{Schematic illustration of Feynman's disk paradox: the electromagnetic angular momentum stored in the field is converted into mechanical rotation of the disk when the current in the solenoid is slowly switched off.}
     \label{fig:feynman_disk}
 \end{figure}
 
A paradigmatic example of this phenomenon is provided by the so-called \emph{Feynman's disk paradox}, discussed by R. P. Feynman in his famous Lectures \cite{feynman_lectures_vol2}. The proposed system consists of an insulating disk, carrying a set of identical point charges uniformly distributed along its rim, and a steady current-carrying solenoid mounted concentrically on the same disk, as shown in Fig. \ref{fig:feynman_disk}. Although the entire apparatus is initially at rest, there is a finite electromagnetic angular momentum \cite{pugh_1967_static_poynting}. Thus, if the current flowing in the solenoid is slowly switched off, through the induced electric field, the angular momentum stored in the electromagnetic field is transferred to the disk, exerting a mechanical torque on it and ultimately causing its rotation. In the whole process, the conservation of the \textit{total angular momentum} is ensured: $\bm L_\text{tot}=\bm{L}_\text{mech}+\bm{L}_\text{em}=\text{\bf const}.$ The quantitative proof of this statement and other related interesting discussions can be found in \cite{corinaldesi_1980_static_angular_momentum,Aguirregabiria_1981,lombardi_feynman_disk_1983,bahder_1985_feynman_paradox, Pantazis_2017_feynman_disk,Jimnez2022TheFP}.

A natural question is whether analogous mechanisms can arise in media whose electromagnetic response differs substantially from that of conventional ones: topological quantum materials, with their low-energy electrodynamics, constitute a particularly interesting class of such media. Among the first experimentally realized 3D topological phases, topological insulators (TIs) \cite{hasan_kane_colloquium_tis,qi_zhang_tis_and_superconductors} are characterized by an insulating bulk together with topologically protected conducting surface states. Unlike ordinary band insulators, a nontrivial topological invariant $\mathbb{Z}_2$ of the occupied electronic bands, through the bulk-boundary correspondence, guarantees the existence of protected conducting surface states.
%
Analogously, Weyl semimetals (WSMs) \cite{yan_felser_2017_review_wsm,hasan_xu_2017_review_wsm,armitage_mele_2018_review_wsm} constitute a 3D topological phase of quantum matter whose low-energy electronic excitations behave as Weyl fermions \cite{weyl_1929_electron_and_gravitation}, namely massless chiral quasiparticles described by the Weyl equation \cite{maggiore_qft}. Their defining feature is the presence of isolated band-touching points, known as \emph{Weyl nodes}, around which the electronic dispersion is linear in all three momentum-space directions. Each Weyl node acts as a monopole of Berry curvature \cite{berry_1983_geometric_phase,xiao_2010_berry_phase_review}, carrying a quantized topological charge identified with the chirality of the corresponding Weyl fermion. Since the total topological charge over the 3D Brillouin zone must vanish, Weyl nodes occur in pairs of opposite chirality (the minimal case being a single pair with zero net chirality), in accordance with the Nielsen-Ninomiya no-go theorem \cite{nielsen_ninomiya_1981_proof1,nielsen_ninomiya_1981_proof2}. The existence of nondegenerate Weyl nodes requires the breaking of either time-reversal or inversion symmetry. Their separation in momentum space is responsible for several unique electromagnetic properties and related to the appearance of topologically protected Fermi-arc surface states \cite{wan_2011_topological_semimetals_and_fermi_arc_surface_states,yang_2011_qhe_in_wsm,burkov_2011_wsm_in_ti_layers,burkov_2011_topological_nodal_semimetals,lv_2015_observation_nodes_taas,xu_2015_observation_of_fermi_arcs_in_na3_bi,huang_2015_negative_magnetoresistance_taas,yan_felser_2017_review_wsm,hasan_xu_2017_review_wsm,armitage_mele_2018_review_wsm,xu_liu_2018_fermiarcs_in_co3sn2,xu_2020_electronic_correlations_in_co3sn2s2,zu_2021_optical_properties_taas_nbas,bonasera_pellegrino_2022_tunable_floquet_wsm}.

Despite their microscopic differences, the electromagnetic response both of TIs and WSMs can be described, in the context of low energy effective field theory, by a so-called \textit{axion term} \cite{wilczek_applications_axion_electrodynamics,hasan_kane_colloquium_tis,qi_zhang_tis_and_superconductors,zyuzin_burkov_2012_wsm_response,vazifeh_2013_em_response_wsm,goswami_2013_3+1_field_teory_wsm,zyuzin_2015_chiral_em_wave_in_wsm,kotov_2018_tunable_nonreciprocity_of_light_in_wsm,sekine_axion_2021_ed_in_topological_materials}. This term, added to the standard electromagnetic Lagrangian density, leads to a set of modified Maxwell's equations, as shown in Sec. \ref{sec:axion_maxwell_equations}, producing modified constitutive relations and novel effects due to the magnetoelectric coupling. In TIs, for example, the axion term gives rise to magnetic responses equivalent to those produced by image magnetic monopoles \cite{qi_zhang_ti_monopole}, produces unconventional Kerr and Faraday rotations in electromagnetic waves \cite{tse_2010_kerr_faraday_effect_tis}, modifies Casimir interactions \cite{grushin_2011_casimir_tis,lopez_2011_casimir_tis} and is responsible for peculiar optical properties \cite{chang_2009_tis-_optical_signature,crosse_2015_electromagnetic_greens_function_layered_tis,franca_2022_modification_of_transition_radiation_3dtis}. In WSMs, the axion response significantly modifies the helicon dispersion compared to conventional metals \cite{pellegrino_2015_helicons_wsm}, affects the bulk and surface plasmons of extended samples \cite{zhou_2015_plasmon_detecting_chirality,song_2017_fermi_arc_plasmons_wsm,andolina_pellegrino_2018_nonlocal_theory_plasmons_wsm,tamaya_2019_surface_plasmons_film_wsm,tsuchikawa_2020_wsm_plasmon_polaritons,bugaiko_2020_surface_plasmons_strained_wsm,heidari_2021_anomalous_plasmons_strained_wsm}, influences the surface plasmon polaritons supported by cylindrical waveguides \cite{peluso_buccheri_2025_wsm_waveguides}, and alters the localized plasmon modes of finite systems \cite{pellegrino_2025_surface_plasmons_wsm}, among several other effects.


In this framework, it is therefore intriguing to investigate how the axion response affects the storage and transfer of electromagnetic angular momentum and whether it can give rise to an analogue of Feynman's disk paradox. In particular, we adopt the geometrical configuration introduced in Ref.~\cite{feynman_paradox_ti}, where the electromagnetic angular momentum generated by a point charge in the vicinity of a spherical TI was investigated. Here, we extend this analysis to a WSM sphere, which, to the best of our knowledge, has not been considered previously in the existing literature. To enable a direct comparison between the two material classes, we also revisit the TI case within the same framework and finally investigate how the stored field angular momentum is converted into mechanical angular momentum.

The paper is structured as follows. In Sec.~\ref{sec:axion_maxwell_equations} we recall axion electrodynamics for TIs and WSMs. In Sec.~\ref{sec:wsm_sphere} we solve the modified Maxwell's equations in the static regime for a WSM sphere in the presence of a point charge, by performing a perturbative expansion in an appropriate parameter associated with the magnetoelectric coupling. Subsequently, the results are compared with the exactly solvable case of a TI sphere. Then, in Sec.~\ref{sec:electromagnetic_angular_momentum_for_a_ti_wsm_sphere}, we calculate the electromagnetic angular momentum for a WSM/TI sphere and a point charge, discussing and comparing the two results.
Finally, in Sec.~\ref{sec:rotation_ti_wsm_sphere}, we study how a quasi-static motion of the point charge generates the rotation of the sphere, deriving the expressions for its angular velocity and rotation angle, and we show that the effect is significantly enhanced in the WSM case. Conclusions are drawn in Sec.~\ref{sec:summary_conclusions}.


\section{Axion Maxwell's equations}\label{sec:axion_maxwell_equations}

To describe the low-energy electromagnetic response of topological materials, such as TIs or WSMs, we add the so-called axion term $\mathcal{L}_\Theta$ to the standard Maxwell Lagrangian density $\mathcal{L}_\text{EM}$ \cite{jackson_classical_electrodynamics}, given by (Gaussian units are used in this work)
\begin{equation}\label{standard_L_EM}
    \mathcal{L}_\text{EM}=c^{-1}\bm{J}\cdot \bm{A}-\rho\varphi+\frac{1}{8\pi}\left(\bm{E}^2-\bm{B}^2\right).
\end{equation}
This axion term $\mathcal{L}_\Theta$ was first introduced in the context of particle physics \cite{wilczek_applications_axion_electrodynamics} and has the following form:
\begin{equation}\label{axion_Lagrangian}
    \mathcal{L}_{\Theta}=\frac{\alpha}{4\pi^2}\Theta(\bm{r},t)\bm{E}\cdot\bm{B}=k\Theta(\bm{r},t)\bm{E}\cdot\bm{B},
\end{equation}
where $\alpha={e^2}/{(\hbar c)}$ is the fine structure constant and $\Theta(\bm{r},t)$ is the so-called \textit{axion field}, which may be in general a function of position and time. For convenience, we have defined $k=\alpha/4\pi^2$. 
The resulting theory is therefore described by the Lagrangian density $\mathcal{L}=\mathcal{L}_\text{EM}+\mathcal{L}_{\Theta}$: the axion term gives rise to a magnetoelectric response, characterized by the \textit{axion polarization} $\bm{P}_\Theta$ and \textit{axion magnetization} $\bm{M}_\Theta$, respectively:
\begin{subequations}
    \begin{align}
        &\bm{P}_\Theta=k\Theta\bm{B},\label{def_axion_polarization}\\
     &\bm{M}_\Theta=k\Theta\bm{E}.\label{def_axion_magnetization}
    \end{align}
\end{subequations}
In linear isotropic media, Maxwell's equations take the form \cite{jackson_classical_electrodynamics}
\begin{subequations}
\begin{align}
    \grad& \cdot \bm D=4\pi\rho_E,\label{div_D_macro_axion}\\
    \grad& \cdot \bm B=0,\label{div_B_macro_axion}\\
     \grad& \times \bm E+{c}^{-1}{\partial_t}\bm B=0,\label{rot_E_macro_axion}\\
     \grad& \times \bm H-{c}^{-1}{\partial_t}\bm D={4\pi}{c}^{-1}\bm J_E,\label{rot_H_macro_axion}
\end{align}
\end{subequations} $\rho_E$ and $\bm J_E$ being the \textit{free} electric charge density and current density, respectively, and we use the auxiliary fields $\bm D$ and $\bm H$ defined as
\begin{subequations}
    \begin{align}
    &\bm D=\bm E+4\pi(\bm P + \bm P_\Theta)=\varepsilon\bm E+4\pi k\Theta \bm B,\label{D_const_relation_axion}\\
    &\bm H={\bm B}-4\pi(\bm M+ \bm M_\Theta)={\mu}^{-1}{\bm B}-4\pi k\Theta\bm E,\label{H_const_relation_axion}
\end{align}
\end{subequations}
where the dimensionless constant quantities $\varepsilon$ and $\mu$ denote the relative dielectric permittivity and the relative magnetic permeability, respectively. Equivalently, Eqs.~(\ref{D_const_relation_axion})-(\ref{H_const_relation_axion}) can be expressed as
\begin{subequations}
    \begin{align}
    \bm D&=\left[\varepsilon+\mu(4\pi k\Theta)^2\right]\bm E +4\pi\mu k\Theta \bm H,\label{D_const_relation_axion_EH}\\
    \bm B&=\mu \bm H+4\pi \mu k\Theta\bm E.\label{B_const_relation_axion_EH}
\end{align}
\end{subequations}
Let us now specify the forms of axion field $\Theta(\bm r, t)$ we are interested in. As for topologically non-trivial insulators preserving either time-reversal or inversion symmetry, one has
\begin{equation}\label{Theta_TI_general}
    \Theta^\text{TI}(\bm r,t)=
    \begin{cases}
        \pi, & \text{topological insulator},\\
        0, & \text{trivial insulator}.
    \end{cases}
\end{equation}
Although $\Theta$ is piecewise constant, its discontinuities at the interfaces give rise to the topological magnetoelectric response. 
Such a description successfully captures the low-energy electromagnetic properties of several experimentally realized TIs, including $\mathrm{Bi_2Te_3}$\cite{zhang_2009_tis_in_bi2se3_bite3_sb2te3}, $\mathrm{Sb_2Te_3}$ \cite{zhang_2009_tis_in_bi2se3_bite3_sb2te3}, $\mathrm{Bi_2Se_3}$ \cite{wu_2016_bi2se3_kerr_faraday} and strained $\mathrm{HgTe}$ \cite{dziom_2017_strained_hgte_magnetoelectric}.

A different situation occurs in WSMs. In the minimal low-energy model \cite{burkov_2011_wsm_in_ti_layers,burkov_2011_topological_nodal_semimetals,armitage_mele_2018_review_wsm}, the Hamiltonian presents two Weyl nodes only:
\begin{equation}\label{hamiltoniana_wsm_2_nodi}
    \hat{H}=\hbar v_{\rm D}\tau^z\bm \sigma\,\cdot(-i\grad+\tau^z\bm b)+\hbar \tau^zb_0.
\end{equation}
Here, $v_{\rm D}$ is the Fermi velocity, $\tau^z$ is the Pauli matrix acting in the node (chirality) space and the 3D vector of
Pauli matrices $\bm \sigma=(\sigma^x,\sigma^y,\sigma^z)^\text{T}$ describes conduction and
valence-band degrees of freedom. The two Weyl
nodes are placed at $\pm \bm b$ and shifted by $2\hbar b_0$ in energy. As shown in Ref. \cite{zyuzin_burkov_2012_wsm_response}, through a unitary transformation, the terms containing $\bm b$ and $b_0$ can be removed from Eq. (\ref{hamiltoniana_wsm_2_nodi}), reducing the Hamiltonian to $\hat{H}=-i\hbar v_{\rm D} \tau^z \bm{\sigma} \cdot \grad.$
However, this transformation simultaneously generates an additional contribution to the electromagnetic action, yielding the axion field
\begin{equation}\label{Theta_WSM_general}
    \Theta^\text{WSM}(\bm r,t)=2\bm b\cdot\bm r-2b_0 t.
\end{equation}
Unlike Eq. (\ref{Theta_TI_general}), the axion field in WSMs depends linearly on space and time. Moreover, its derivatives, directly determined by $\bm b$ and $b_0$, enter the axion Maxwell's equations, making the magnetoelectric response a bulk property rather than only an interfacial effect, in contrast with TIs. In this work, we focus on Eq. (\ref{Theta_WSM_general}) with $b_0=0$, such that the inversion
symmetry is conserved. We consider a finite $\bm b$
related to the breaking of the time-reversal symmetry. This phase has been realized or predicted in different
materials, such as $\text{EuCd}_2\text{As}_2$ \cite{wang_2019_wsm_eucd2as2,wang_2016_anisotropic_transport_eucd2as2,krishna_2018_1st_principles_of_eucd2as2}, $\text{HgCr}_2\text{Se}_4$ \cite{xu_2011_wsm_hgcr2se4}, $\text{MnBi}_2\text{Te}_4$ \cite{li_2019_ti_mnbi2te4},
$\text{MnSn}_2\text{Sb}_2\text{Te}_6$ \cite{gao_2023_wsm_mnx2b2t6,boulton_2024_searching_antiferromagnetic_wsm}, KCrTe \cite{liu_2024_wsm_kcrte_rbcrte}, RbCrTe \cite{liu_2024_wsm_kcrte_rbcrte}, $\text{Eu}_2\text{Ir}_2\text{O}_7$ \cite{sushkov_2015_wsm_eu2ir2o7}, $\text{Mn}_3\text{Sn}$ \cite{cao_2023_wsm_mn3sn}, $\text{Co}_3\text{Sn}_2\text{S}_2$ \cite{iohani_2023_wsm_co3snsS2}.

\section{WSM sphere in the presence of a point charge}\label{sec:wsm_sphere}

\begin{figure}[t]
   \includegraphics[width=0.75\columnwidth]{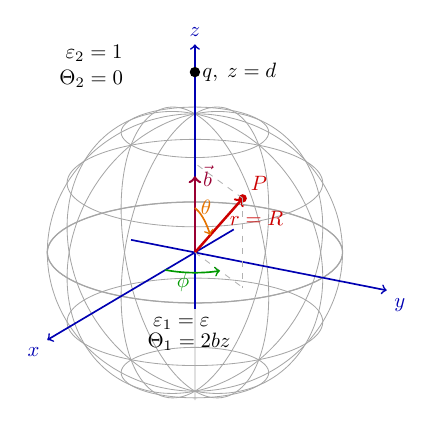}
  \caption{\label{fig:disegno_sfera_wsm}%
   Geometry of the system. A WSM sphere of radius $R$, centered at the origin, is characterized by relative dielectric permittivity $\varepsilon$ and axion field $\Theta=2bz$, and is surrounded by vacuum. A point charge $q=Ne$ is placed on the symmetry axis at $(0,0,d)$, with $d>R$. The point $P$ denotes a generic point inside the sphere.
   }
\end{figure}
 
We begin by analyzing the electrostatic problem of a WSM sphere of radius $R$, centered at the origin and surrounded by vacuum, subjected to a point charge $q=Ne$, $N\in\mathbb{Z}$, located at $\mathbf{r}_0=d\hat{\mathbf z}$, with $d>R$ (see Fig.~\ref{fig:disegno_sfera_wsm}). We assume that the  Weyl-node separation vector $\bm{b}$, for simplicity, is given by $\bm{b}=b\hat{\bm z}$, $b>0$, to preserve azimuthal symmetry, and $b_0=0$, to preserve inversion symmetry. As far as the ordinary polarization and magnetization are concerned, the sphere is a linear, isotropic, and homogeneous medium with relative static dielectric permittivity $\varepsilon$ and relative magnetic permeability $\mu\simeq1$. Hence, recalling Eq. (\ref{Theta_WSM_general}), in the present situation, one has $\mu(\bm r)=1$ and
\begin{subequations}
    \begin{align}
    \varepsilon(\mathbf r)=\varepsilon\,H(R-r)+H(r-R)=\begin{cases}
    \varepsilon, & r<R,\\
    1, & r>R,
  \end{cases}\label{eq:varepsilon_sphere}\\
    \Theta(\mathbf r)=2bzH(R-r)=
  \begin{cases}
    2bz, & r<R,\\
    0, & r>R,
  \end{cases}
  \label{eq:ThetaWSM_sphere}
\end{align}
\end{subequations}
$H(x)$ denoting the Heaviside step function. In the static regime, Eqs.~(\ref{div_D_macro_axion})-(\ref{rot_H_macro_axion}) together with
the constitutive relations (\ref{D_const_relation_axion_EH})-(\ref{B_const_relation_axion_EH}) read as follows
\begin{subequations}
    \begin{align}
    \grad&\cdot\mathbf D = 4\pi q\,\delta(\mathbf r-\mathbf r_0),\label{div_D_WSM_static}\\
    \grad&\cdot\mathbf B = 0,\label{div_B_WSM_static}\\
    \grad&\times\mathbf E = 0,\label{rot_E_WSM_static}\\
    \grad&\times\mathbf H = 0,\label{rot_H_WSM_static}
\end{align}
\end{subequations}
with $\varepsilon(\bm r)$, $\mu(\bm r)$ and $\Theta(\bm r)$ specified as above. It is possible to introduce electric and magnetic scalar potentials $\varphi$ and $\psi$ such that $\mathbf E=-\grad\varphi$ and $\mathbf H=-\grad\psi$.
Using these two potentials, one can solve Eqs.~(\ref{div_D_WSM_static})-(\ref{div_B_WSM_static}) restricting them in the two domains $0<r<R$ (domain 1) and $r>R$ (domain 2), matching then the solutions at $r=R$ through the conditions \cite{jackson_classical_electrodynamics}
\begin{subequations}
    \begin{align}
    &\hat{\mathbf r}\times\left[\mathbf E_1-\mathbf E_2\right]_{r=R}=0,\label{TI_match_E}\\
  &\hat{\mathbf r}\times[\mathbf H_1-\mathbf H_2]_{r=R}=0,\label{TI_match_H}\\
  &\hat{\mathbf r}\cdot[\mathbf D_1-\mathbf D_2]_{r=R}=0,\label{TI_match_D}\\
  &\hat{\mathbf r}\cdot[\mathbf B_1-\mathbf B_2]_{r=R}=0.\label{TI_match_B}
\end{align}
\end{subequations}
In the two regions, Eqs.~(\ref{div_D_WSM_static})-(\ref{div_B_WSM_static}) become
\begin{subequations}
    \begin{empheq}[left=\empheqlbrace]{align}
    \nabla^{2}&\varphi_{1}
      +\tfrac{\lambda}{\varepsilon}\,\partial_{z}\psi_{1}
      +\tfrac{\lambda^{2}}{\varepsilon}\,z\,\partial_{z}\varphi_{1}
      =0,\label{eq_phi1_eq}\\
    \nabla^{2}&\psi_{1}
      +\lambda\,\partial_{z}\varphi_{1}
      -\tfrac{\lambda^{2}}{\varepsilon}\,z\,\partial_{z}\psi_{1}
      -\tfrac{\lambda^{3}}{\varepsilon}\,z^{2}\,\partial_{z}\varphi_{1}
      =0,\label{eq_psi1_eq}
      \end{empheq}
\end{subequations}
for $0<r<R$, where we have defined the inverse length $\lambda=8\pi kb$, and
\begin{subequations}\label{eq_potentials_domain2_C}
\begin{empheq}[left=\empheqlbrace]{align}
    \nabla^{2}\varphi_{2} &=-4\pi q \delta(\bm{r}-d\hat{\bm z}),\label{eq_phi2_eq}\\
    \nabla^{2}\psi_{2} &=0,\label{eq_psi2_eq}
\end{empheq}
\end{subequations}
for $r>R$. The general solution to Eqs.~(\ref{eq_phi2_eq})-(\ref{eq_psi2_eq}) is known \cite{jackson_classical_electrodynamics}, whereas Eqs.~(\ref{eq_phi1_eq})-(\ref{eq_psi1_eq}) are coupled linear PDEs with non constant coefficients. Employing ansatzes of the form \cite{jackson_classical_electrodynamics} $ \varphi_1(r,\theta)=\sum_{n=0}^\infty a_n(r)P_n(\cos\theta)$ and $\psi_1(r,\theta)=\sum_{n=0}^\infty A_n(r)P_n(\cos\theta),$ $P_n(x)$ being the $n$-th Legendre polynomial \cite{jackson_classical_electrodynamics,gradshteyn2007} (see Appendix~\ref{sec:legendre_polynomials} for useful identities employed throughout the calculations), we find an infinite system of coupled ODEs for $a_n(r)$ and $A_n(r)$ with bandwidth equal to 7, i.e. coefficients of order $n$ are coupled to coefficients of orders $n\pm3$, $n\pm2$, $n\pm1$. Such a system would not be analytically solvable unless it were truncated to a finite number of harmonics, losing mathematical and physical information.

\subsection{Perturbative scheme}
\label{sec:WSM_scheme}
 
It is useful to rescale all lengths to the radius of the sphere: $\boldsymbol\rho={\mathbf r}/{R}=(u,v,w)$, $\beta=bR,$ and define the following dimensionless measure of the magnetoelectric coupling:
\begin{equation}
    \eta = \lambda R=8\pi k bR=\frac{2\alpha\beta}{\pi}.
\end{equation}
With these definitions, Eqs.~(\ref{eq_phi1_eq})-(\ref{eq_psi1_eq}) read
\begin{subequations} 
    \begin{empheq}[left=\empheqlbrace]{align}
    \hspace{-0.3mm}\nabla_\rho^2&\varphi_1+\tfrac{\eta}{\varepsilon}\,\partial_w\psi_1
      +\tfrac{\eta^2}{\varepsilon}\,w\,\partial_w\varphi_1=0,\label{eq_phi1_eq_rho}\\
    \hspace{-0.3mm}\nabla_\rho^2&\psi_1+\eta\,\partial_w\varphi_1
      -\tfrac{\eta^2}{\varepsilon}\,w\,\partial_w\psi_1
      -\tfrac{\eta^3}{\varepsilon}\,w^2\,\partial_w\varphi_1=0,\label{eq_psi1_eq_rho}
      \end{empheq}
\end{subequations}
for $0<\rho<1$. For typical WSMs, one has $\varepsilon\sim5-20$ \cite{yan_felser_2017_review_wsm,hasan_xu_2017_review_wsm,armitage_mele_2018_review_wsm,xu_2020_electronic_correlations_in_co3sn2s2,zu_2021_optical_properties_taas_nbas,martin_ruiz_2019_charge_near_wsm} and
$2b\sim0.01-0.1\text{\AA}^{-1}$ \cite{yan_felser_2017_review_wsm,hasan_xu_2017_review_wsm,armitage_mele_2018_review_wsm}, which for a nanosphere, $R\sim10^2\,\text{nm}$, gives $2\beta\sim10-100$. Hence, $\eta$ may be regarded as a small quantity: $\eta\sim0.02-0.2$. 
This estimate justifies the adoption of a controlled perturbative approach while ensuring that the leading-order magnetoelectric corrections are accurately captured.
Since $0<w<1$, we may therefore treat the
terms carrying increasing powers of $\eta$ in Eqs.~(\ref{eq_phi1_eq_rho})-(\ref{eq_psi1_eq_rho}) as perturbations of the Laplacian. 
So, we seek solutions of the following forms
\begin{subequations}\label{eq:WSM_expansion}
\begin{align}
  \varphi_1&=\sum_{j=0}^\infty \eta^{\,j}\varphi_1^{(j)},\\
  \psi_1&=\sum_{j=0}^\infty \eta^{\,j}\psi_1^{(j)}.
\end{align}
\end{subequations}
\noindent For $\eta=0$ the sphere behaves as an ordinary dielectric such that $\psi=0$ in the whole space without a source of
magnetic field. Thus, we set $\psi_1^{(0)}=0$.
Rewriting Eqs.~(\ref{eq_phi1_eq_rho})-(\ref{eq_psi1_eq_rho}) order by order, one verifies that the electric potential acquires genuinely new sources generated by the magnetoelectric coupling $\eta$ only at \emph{even} orders in
$\eta$, and the magnetic potential only at \emph{odd} orders. Therefore, the remaining terms represent arbitrary harmonic functions and can be set equal to zero, and one has
\begin{subequations}\label{eq:WSM_parity}
\begin{align}
  \varphi_1&=\varphi_1^{(0)}+\eta^2\varphi_1^{(2)}+\mathcal{O}(\eta^4),\\
  \psi_1&=\eta\,\psi_1^{(1)}+\eta^3\psi_1^{(3)}+\mathcal{O}(\eta^5).
\end{align}
\end{subequations}
Since we are interested in the leading magnetoelectric response, we work to first order in $\eta$, where the hierarchy reduces to
\begin{subequations}\label{eq_potentials_domain1_C}
\begin{empheq}[left=\empheqlbrace]{align}
    \nabla^{2}&\varphi^{(0)}_{1}=0,\label{eq_phi1_eq_1storder}\\
    \nabla^{2}&\psi^{(1)}_{1} +\tfrac{1}{R}\,\partial_z\varphi_1^{(0)}=0,\label{eq_psi1_eq_1storder}
\end{empheq}
\end{subequations}
having switched back to the original unscaled variable $0<r<R$.
 
\subsection{First-order solution}
\label{sec:WSM_solution}
 
The electric potential $\varphi_1^{(0)}$ is harmonic and regular at the
origin~\cite{jackson_classical_electrodynamics}, hence one has 
\begin{equation}
  \varphi_1(r,\theta)=q\sum_{n=0}^{\infty}a_n\,r^n P_n(\cos\theta)+\mathcal{O}(\eta^2), \qquad 0<r<R.
  \label{eq:WSM_phi_in}
\end{equation}
Its source term for $\psi_1^{(1)}$ can be recast, using Eq. (\ref{partial_z_r^n_P_n}), contained in Appendix~\ref{sec:legendre_polynomials}, as
\begin{equation}
  \partial_z\varphi_1^{(0)}
    =q\sum_{n=0}^{\infty}(n+1)\,a_{n+1}\,r^n P_n(\cos\theta).
\end{equation}
Looking for a solution $\psi_1^{(1)}=q\sum_{n=0}^\infty F_n(r)P_n(\cos\theta)$, each
$F_n$ obeys an ODE whose particular solution is $F_n(r)=C_n r^{n+2}$, with $C_n=-\tfrac{n+1}{2(2n+3)R}\,a_{n+1}.$
Thus, adding the regular harmonic solution $A_n r^n$, the internal magnetic
potential, for $0<r<R$, reads
\begin{equation}
  \psi_1(r,\theta)=\eta\,q\sum_{n=0}^{\infty}
     \bigl(A_n r^n+C_n r^{n+2}\bigr)P_n(\cos\theta)+\mathcal{O}(\eta^3).
  \label{eq:WSM_psi_in}
\end{equation}
The term $C_nr^{n+2}$ is \emph{non-harmonic}: it is the hallmark of the
spatially varying axion field expressed in Eq.~(\ref{Theta_WSM_general}), whose non-zero
bulk gradient acts as an effective source. Outside the sphere, for $r>R$, imposing regularity at infinity \cite{jackson_classical_electrodynamics}, one has 
\begin{align}
  \varphi_2(r,\theta)&=\frac{q}{|\mathbf r-d\hat{\mathbf z}|}
      +q\sum_{n=0}^{\infty}b_n\,r^{-(n+1)}P_n(\cos\theta)+\mathcal{O}(\eta^2),
  \label{eq:WSM_phi_out}\\
  \psi_2(r,\theta)&=\eta\,q\sum_{n=0}^{\infty}
      B_n\,r^{-(n+1)}P_n(\cos\theta)+\mathcal{O}(\eta^3).
  \label{eq:WSM_psi_out}
\end{align}
 Imposing the matching conditions shown in Eqs.~(\ref{TI_match_E})-(\ref{TI_match_B}) on
Eqs.~(\ref{eq:WSM_phi_in}) and (\ref{eq:WSM_psi_in})-(\ref{eq:WSM_psi_out}) gives a $4\times4$ linear system for 
$a_n,\, b_n,\, A_n,$ and $B_n$. Moreover, Eq.~(\ref{Legendre_espansione_1/r}) has been used for the Coulomb term in $\varphi_2$. 
Because we are considering only terms up to first order in $\eta$, the
matching condition which involves $\bm D$ consistently reduces to that of an ordinary
dielectric sphere, the magnetoelectric corrections being of order
$\eta^2$. Without this simplification, one would face an infinite
tridiagonal system in $a_n,a_{n\pm2}$, which is not solvable in closed form.
Solving the system, one has the following
\begin{subequations}
\begin{align}
  a_n&=\frac{2n+1}{\varepsilon_n}\frac{1}{d^{\,n+1}},
  \label{eq:aWSM}\\
  b_n&=\frac{n(1-\varepsilon)}{\varepsilon_n}\frac{R^{2n+1}}{d^{\,n+1}},
  \label{eq:bWSM}\\
  A_n&=\frac{1}{2n+1}\!\left[\frac{n+1}{2\varepsilon_{n+1}}\frac{R^2}{d^2}
       -\frac{n(n-1)}{\varepsilon_{n-1}}\right]\!\frac{1}{Rd^{n}},
  \label{eq:AWSM}\\
  B_n&=-\frac{n}{2n+1}\left[\frac{n+1}{\varepsilon_{n+1}}\frac{R^2}{d^2}
       +\frac{n-1}{\varepsilon_{n-1}}\right]\!\frac{R^{2n}}{d^{n}},
  \label{eq:BWSM}\\
  C_n&=-\frac{n+1}{2\varepsilon_{n+1}}\frac{1}{Rd^{n+2}},
  \label{eq:CWSM}
\end{align}
\end{subequations}
where we have defined 
\begin{equation}
\varepsilon_n=1+n(1+\varepsilon)~.
\end{equation}
Equivalently, the magnetic coefficients $A_n$ and $B_n$ can be written directly as
combinations of electric ones of adjacent order:
\begin{subequations}
\begin{align}
  &A_n=\tfrac{n+1}{2(2n+1)(2n+3)}Ra_{n+1}
      -\tfrac{n(n-1)}{(2n-1)(2n+1)R}{a_{n-1}},
  \label{A_n_combinazione_lineare_di_a_npm1}\\
  &B_n=-\tfrac{n(n+1)}{(2n+1)(2n+3)}R^{2n+2}a_{n+1}
      -\tfrac{n(n-1)}{(2n-1)(2n+1)}R^{2n}a_{n-1}.
  \label{B_n_combinazione_lineare_di_a_npm1}
\end{align}
  \label{AB_n_combinazione_lineare_di_a_npm1}
\end{subequations}
\subsection{Comparison with the TI case}
\label{sec:WSM_vs_TI}

The solution of the problem of a TI sphere was first presented in \cite{lindell_biisotropic_sphere}, where the electrostatic problem of a bi-isotropic sphere in the presence of a point charge is examined. The TI case is simply obtained with the substitutions $\varepsilon\to\varepsilon+(4\pi k\Theta)^2$, $\mu\to 1$, $\xi \to 4\pi k\Theta$ and $\zeta \to 4\pi k\Theta$ in Equations (1)-(2) of  \cite{lindell_biisotropic_sphere}, where $\Theta$ is given by Eq. (\ref{Theta_TI_general}). 
An equivalent result is also presented in the Supplementary Material of Ref.~\cite{qi_zhang_ti_monopole}.
Referring to scalar potentials $\varphi$ and $\psi$ of the forms shown in Eqs. (\ref{eq:WSM_phi_in}) and (\ref{eq:WSM_psi_in})-(\ref{eq:WSM_psi_out}), replacing $2\beta$ with $\Theta$ in the prefactors of $\psi$, the coefficients $a_n$, $b_n$, $A_n$, $B_n$, $C_n$ in the case of a TI sphere are given by the following expressions:
\begin{subequations}\label{abABC_n_TI_esatto}
\begin{align}
    a^\text{TI}_n&=\frac{2n+1}{\varepsilon_n[1+\tfrac{n(n+1)}{\varepsilon_n(2n+1)}(4\pi k\Theta)^2]}\frac{1}{d^{n+1}},\label{a_n_TI_esatto}\\
    b^\text{TI}_n&=\frac{n(1-\varepsilon)}{\varepsilon_n}\frac{1-\tfrac{n+1}{(2n+1)(1-\varepsilon)}(4\pi k\Theta)^2}{1+\tfrac{n(n+1)}{\varepsilon_n(2n+1)}(4\pi k\Theta)^2}\frac{R^{2n+1}}{d^{n+1}},\label{b_n_TI_esatto}\\
    A^\text{TI}_n&=B^\text{TI}_n=-\frac{n}{\varepsilon_n[1+\tfrac{n(n+1)}{\varepsilon_n(2n+1)}(4\pi k\Theta)^2]}\frac{R^{2n+1}}{d^{n+1}},\label{A_n_B_n_TI_esatto}\\
     C^\text{TI}_n&=0.\label{C_n_TI_esatto}
\end{align}
\end{subequations}
Let us point out that a different sign in Eq.~(\ref{A_n_B_n_TI_esatto}), reported by some authors, is due to the different convention on the sign of the axion Lagrangian (\ref{axion_Lagrangian}).
It is worth observing that Eq.~(\ref{abABC_n_TI_esatto}) may be suitably expanded in powers of $4\pi k\Theta=\alpha\Theta/\pi=\alpha$, analogously to what we have done in the WSM case using the parameter $\eta$. Retaining the leading order terms, we obtain the following
\begin{subequations}
\begin{align}
    &a^\text{TI}_n=\frac{2n+1}{\varepsilon_n}\frac{1}{d^{n+1}}+\mathcal{O}(\alpha^2),\label{a_n_TI_1storder}\\
    &b^\text{TI}_n=\frac{n(1-\varepsilon)}{\varepsilon_n}\frac{R^{2n+1}}{d^{n+1}}+\mathcal{O}(\alpha^2),\label{b_n_TI_1storder}\\
    &A^\text{TI}_n=B^\text{TI}_n=-\frac{n}{\varepsilon_n} \frac{R^{2n+1}}{d^{n+1}}+\mathcal{O}(\alpha^2).\label{A_n_B_n_TI_1storder}
\end{align}
\end{subequations}
By comparing Eqs.~(\ref{eq:aWSM})-(\ref{eq:bWSM}) with the leading order TI coefficients in Eqs.~(\ref{a_n_TI_1storder})-(\ref{b_n_TI_1storder}), we see that the
electric coefficients $a_n$ and $b_n$ are identical in the two
cases. At this order in perturbation theory, both spheres respond to the
electric potential exactly as an ordinary dielectric. In contrast, the magnetic coefficients $A_n$ and $B_n$ differ significantly in the two situations. Eq.~(\ref{AB_n_combinazione_lineare_di_a_npm1}) couples electric harmonics of orders $n \pm 1$, due to the axion field linearly varying in position, see Eq.~\eqref{Theta_WSM_general}. Conversely, the form of Eq. (\ref{A_n_B_n_TI_1storder}) involves only the electric harmonics of order $n$, since it is $A^\text{TI}_n=B^\text{TI}_n=-\frac{n}{2n+1}R^{2n+1}a_n^\text{TI}$, even in the non-perturbative case described by Eq. (\ref{A_n_B_n_TI_esatto}). To conclude, we observe that Eq. (\ref{C_n_TI_esatto}) shows that non-harmonic terms are absent in the TI case. In contrast to the WSM case, where Eq. (\ref{eq:CWSM}) is valid, the axion field in Eq.~(\ref{Theta_TI_general}) is piecewise constant and does not modify Maxwell's equations in the bulk, only modifying the matching conditions at the interface.

%
\section{Electromagnetic angular momentum for a WSM/TI sphere}\label{sec:electromagnetic_angular_momentum_for_a_ti_wsm_sphere}

\begin{figure*}[t]
  \centering
  \begin{overpic}[width=0.48\textwidth]{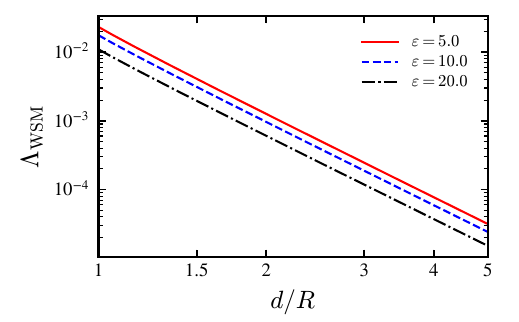}%
    \panellab{(a)}%
  \end{overpic}\hfill
  \begin{overpic}[width=0.48\textwidth]{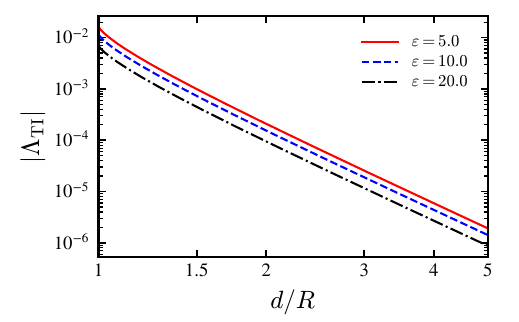}%
    \panellab{(b)}%
  \end{overpic}
 
  \vspace{4pt}
 
  \begin{overpic}[width=0.48\textwidth]{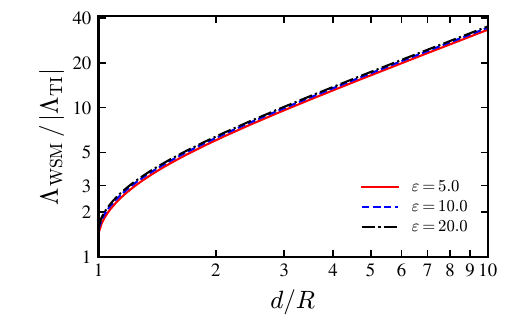}%
    \panellab{(c)}%
  \end{overpic}\hfill
  \begin{overpic}[width=0.48\textwidth]{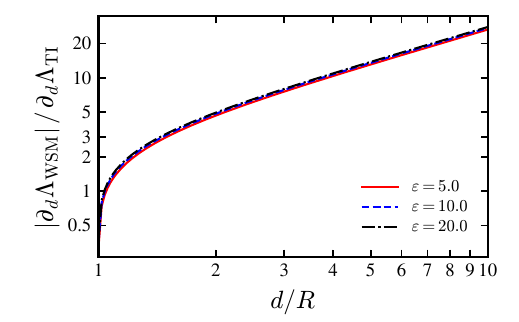}%
    \panellab{(d)}%
  \end{overpic}
 
  \caption{
(a) Dimensionless factor $\Lambda_{\rm WSM}$, defined in Eq.~\eqref{eq:Lambda_WSM}, as a function of $d/R$. (b) Corresponding quantity $|\Lambda_{\rm TI}|$, defined in Eq.~\eqref{eq:Lambda_TI}. (c) Ratio $\Lambda_{\rm WSM}/|\Lambda_{\rm TI}|$, showing the enhancement of the electromagnetic angular momentum in the WSM case with respect to the TI case. (d) Ratio $|\partial_d\Lambda_{\rm WSM}|/\partial_d\Lambda_{\rm TI}$, relevant for the mechanical torque discussed in Sec.~\ref{sec:rotation_ti_wsm_sphere}. All plots are shown on logarithmic scales for relative dielectric permittivities $\varepsilon=5,10,20$.
    }
  \label{fig:panels}
\end{figure*}

Starting from electric and magnetic scalar potentials expressed in Eqs.~(\ref{eq:WSM_phi_in}) and (\ref{eq:WSM_psi_in})-(\ref{eq:WSM_psi_out}), we evaluate the angular momentum stored in the electromagnetic field for the WSM and TI configurations. 
We adopt the Abraham prescription, in which the momentum density of the field in a medium is $\mathbf p^{\mathrm em}=\mathbf S/c^2=(1/4\pi c)\,\mathbf E\times\mathbf H$, $\mathbf S$ being the Poynting vector \cite{jackson_classical_electrodynamics, brevik_1979_abraham_minkowski, barnett_2010_enigma_optical_momentum}. The associated angular momentum density is $\mathbf l^{\mathrm{em}}=\mathbf r\times\mathbf p^{\mathrm{em}}
    =({1}/{4\pi c})\,\mathbf r\times(\mathbf E\times\mathbf H)$.
Writing $\mathbf E=E_r\hat{\mathbf r}+E_\theta\hat{\boldsymbol\theta}$ and
$\mathbf H=H_r\hat{\mathbf r}+H_\theta\hat{\boldsymbol\theta}$ (both fields
are $\phi$-independent due to azimuthal symmetry), one has
$\mathbf l^{\mathrm{em}}=-(r/4\pi c)(E_rH_\theta-E_\theta H_r)
\hat{\boldsymbol\theta}$. However, because
$\hat{\boldsymbol\theta}=\cos\theta\cos\phi\,\hat{\mathbf x}
+\cos\theta\sin\phi\,\hat{\mathbf y}-\sin\theta\,\hat{\mathbf z}$, the
$x$ and $y$ components vanish upon integration over $\phi$, so only the
$z$ component of the total angular momentum remains
\begin{equation}
  \mathbf L^{\mathrm{em}}=L_z^{\mathrm{em}}\hat{\mathbf z}
    =\frac{1}{4\pi c}\int_{\mathbb R^3}\!\!d^3\bm r\;
      r\sin\theta(E_rH_\theta-E_\theta H_r)\hat{\mathbf z}.
  \label{eq:Lz_integral}
\end{equation}
It is convenient to separate the radial integration into the three regions $0<r<R$, $R<r<d$ and $r>d$: the calculation together with some intermediate results is reported in Appendix~\ref{sec:calculation_em_appendix}. After some algebraic manipulations, one reaches the following compact result
\begin{subequations}\label{eq:Lz_central}
\begin{align}
  \mathbf L^{\mathrm{em}}&=L^{\mathrm{em}}_z\hat{\bm z}=\frac{\kappa}{\pi}N^2\alpha^2\hbar\,\Lambda\, \hat{\mathbf z}, \\
  \kappa&=
  \begin{cases}
    2\beta=2bR, & \text{WSM},\\ \Theta=\pi, & \text{TI},
  \end{cases} 
 \end{align} 
\end{subequations} 
\noindent which shows that the field angular momentum scales
quadratically with the point charge (i.e. with $N^2$, since $q=Ne$), with
$\alpha^2$, and with the dimensionless factor $\kappa/\pi$, which encodes
the topological character of the medium. The remaining dimensionless factor $\Lambda$, which differs for the two material classes, is given by the sum of three contributions $\Lambda_i$, for $i=1,2,3$, originating respectively from the
internal region, from the intermediate one, and from the external one. They
depend only on the multipole coefficients of the potentials $\varphi$ and
$\psi$, and their general expressions are collected in Eqs.~(\ref{eq:Lambda1_gen})-(\ref{eq:Lambda3_gen}) of Appendix~\ref{sec:calculation_em_appendix}.

We now focus on $\Lambda$ for the two material classes. In the WSM case, it takes the following form:
\begin{widetext}
\begin{equation}
  \Lambda_{\mathrm{WSM}}(\varepsilon,R/d)
    =(\varepsilon-1)\sum_{n=0}^{\infty}
     \frac{2(n+1)^2(2n+3)}{(2n+1)^2(2n+5)^2\,\varepsilon_{n+1}^2}
     \left(\frac{R}{d}\right)^{2n+4}+\mathcal{O}(\eta^2),
  \label{eq:Lambda_WSM}
\end{equation}
\end{widetext}
whose dependence on $d/R$ is shown in Fig.~\ref{fig:panels}~(a). Two features
are noteworthy. First, $\Lambda_{\mathrm{WSM}}$ contains only \emph{even}
powers of $R/d$. Second, it is identically zero when $\varepsilon=1$. 
If the sphere possessed only the axion-induced polarization, without ordinary dielectric response, i.e., $\varepsilon=1$, then the angular momentum of the field would not be stored at the lowest order in $\eta$.
For a charge far from the sphere, $R/d\ll1$, the term labeled $n=0$ dominates and
\begin{equation}
  \Lambda_{\mathrm{WSM}}^{\infty}\approx
    \frac{6(\varepsilon-1)}{25(\varepsilon+2)^2}\left(\frac{R}{d}\right)^{4}.
  \label{eq:Lambda_WSM_asy}
\end{equation}
\begin{widetext}
Analogously, if one specifies $\Lambda$ to the TI case, the following result is obtained:
\begin{equation}
  \Lambda_{\mathrm{TI}}(\varepsilon,R/d)
    =-(\varepsilon-1)\sum_{n=0}^{\infty}
     \frac{(n+1)(n+2)}{(2n+3)(2n+5)\,\varepsilon_{n+1}\varepsilon_{n+2}}
     \left(\frac{R}{d}\right)^{2n+5}+\mathcal{O}(\alpha^2),
  \label{eq:Lambda_TI}
\end{equation}
\end{widetext}
whose dependence on $d/R$ is shown in Fig.~\ref{fig:panels}~(b). This result
differs from the one reported in Ref.~\cite{feynman_paradox_ti}: there, the
internal contribution $\Lambda_1$, shown in Eq.~(\ref{eq:Lambda1_gen}) of
Appendix~\ref{sec:calculation_em_appendix}, arising from the cross term
$A_n a_{n+1}-A_{n+1}a_n$ and not vanishing [cf.\
Eqs.~(\ref{a_n_TI_1storder}) and (\ref{A_n_B_n_TI_1storder})], was omitted.
For $R/d\ll1$, the dominant asymptotic behavior is
\begin{equation}
  \Lambda_{\mathrm{TI}}^{\infty}\approx
    -\frac{2(\varepsilon-1)}{15(\varepsilon+2)(2\varepsilon+3)}
     \left(\frac{R}{d}\right)^{5},
  \label{eq:Lambda_TI_asy}
\end{equation}
in contrast with the $(R/d)^3$ scaling reported in
Ref.~\cite{feynman_paradox_ti}. 
Analogously to the WSM case, $\Lambda_{\mathrm{TI}}$ vanishes for $\varepsilon=1$ at this order of perturbation theory. Unlike the WSM case, it contains only \emph{odd} powers of $R/d$.
 
Comparing Eqs.~\eqref{eq:Lambda_WSM} and \eqref{eq:Lambda_TI}, the two field
angular momenta have opposite sign and different magnitude. The positive ratio
$\Lambda_{\mathrm{WSM}}/|\Lambda_{\mathrm{TI}}|$ is larger than unity for
every value of $d/R$ and increases as the charge is moved away, as shown in
Fig.~\ref{fig:panels}~(c), consistent with the asymptotic
forms shown in Eqs.~\eqref{eq:Lambda_WSM_asy} and \eqref{eq:Lambda_TI_asy}.
 
For the analysis of the mechanical torque in
Sec.~\ref{sec:rotation_ti_wsm_sphere}, we also show that the positive ratio
$|\partial_d\Lambda_{\mathrm{WSM}}|/\partial_d\Lambda_{\mathrm{TI}}$ [cf. Eqs. (\ref{eq:dLambda_WSM})-(\ref{eq:dLambda_TI}) of Appendix~\ref{sec:calculation_em_appendix}] is larger than unity for essentially any value of $d/R$ and increases linearly as the point charge is moved away from the sphere, as shown in Fig.~\ref{fig:panels} (d) and as can be inferred from
Eqs.~(\ref{eq:Lambda_WSM_asy}) and (\ref{eq:Lambda_TI_asy}):
\begin{equation}
  \frac{|\partial_d\Lambda_{\mathrm{WSM}}^{\infty}|}
       {\partial_d\Lambda_{\mathrm{TI}}^{\infty}}
    \approx\frac{36}{25}\,\frac{2\varepsilon+3}{\varepsilon+2}\,\frac{d}{R}.
  \label{eq:dLambda_ratio}
\end{equation}
In all cases, the WSM shows a significantly larger response than the TI, a feature that will directly impact the rotation of the sphere discussed in the following section.


\section{Rotation of the sphere}\label{sec:rotation_ti_wsm_sphere}

The static configuration analyzed so far is characterized by a non-zero angular momentum of the electromagnetic field, given by Eq.~\eqref{eq:Lz_central}. We now show that, as in Feynman's disk paradox~\cite{feynman_lectures_vol2}, a mechanical torque arises when this field angular momentum is varied in time, and we estimate the resulting rotation of the sphere.
Let the point charge be displaced along the symmetry axis from $d_0$ at $t=0$ to $d_1$ at $t=t_1$, at constant velocity:
\begin{subequations}\label{eq:d_of_t}
\begin{align}
  d(t)&=d_0+\frac{\Delta d}{t_1}\,t,\\
  \Delta d&=d_1-d_0,\\
    \dot d&=\frac{\Delta d}{t_1}~,
\end{align}  
\end{subequations}
\noindent with the center of the sphere held fixed, and where $\dot d$ represents the time derivative of the coordinate $d(t)$.
We take $t_1$ large, such that the
motion is quasi-static and radiative effects can be neglected. In the
Abraham framework \cite{brevik_1979_abraham_minkowski}, assuming that the fields decay sufficiently rapidly at infinity, the conservation of the total angular momentum in
$\mathbb R^3$
reads
\begin{subequations}\label{eq:angmom_conservation}
\begin{align}
  &\frac{d}{dt}\int_{\mathbb R^3}\!\!d^3\bm r\,\mathbf l^{\mathrm{em}}
   +\boldsymbol\tau^{\mathrm{matter}}=0,
  \\
  &\boldsymbol\tau^{\mathrm{matter}}=\int_{\mathbb R^3}\!\!d^3\bm r\,
    \mathbf r\times\mathbf f,
\end{align}\end{subequations}
\noindent where $\mathbf f$ denotes the Abraham force density \cite{brevik_1979_abraham_minkowski}, and
$\boldsymbol\tau^{\mathrm{matter}}$ represents the total mechanical torque acting on the matter.
Since $\dot{d}$ is small, we can perform an adiabatic expansion of our fields:
\begin{equation}
    \bm E(t)=\bm E_0(d(t))+\bm E_1(d(t))+\mathcal{O}(\dot{d}^2),
\end{equation}
and similarly for
$\mathbf H$, where the subscript $0$ denotes the static fields generated by the potentials of Sec. \ref{sec:wsm_sphere} and $\mathbf E_1$ and $\mathbf H_1$ are terms of order $\mathcal{O}(\dot d)$. These higher order terms are determined by solving order by order with respect to $\dot{d}$ the following equations:
\begin{subequations}\label{eq:adiabatic}
\begin{align}
        &\grad \cdot\bm D=4\pi q\delta(\bm r-d(t)\hat{\bm z}),\label{adiabatic_div_D}\\
        &\grad \cdot \bm B=0,\label{adiabatic_div_B}\\
        &\grad\times\bm E+{c}^{-1}{\partial_t}\bm B=0,\label{adiabatic_rot_E}\\
        &\grad \times \bm H-{c}^{-1}{\partial_t}\bm D={4\pi}{c}^{-1}q\dot{d}\delta(\bm r-d(t)\hat{\bm z})\,\hat{\bm z},\label{adiabatic_rot_H}
\end{align}
\end{subequations}
which extend the static case described by Eqs.~(\ref{div_D_WSM_static})-(\ref{rot_H_WSM_static}). Since the fields depend on time only through $d(t)$, we may write ${d \star}/{dt}=\dot{d}\,{\partial}\star/\partial d$; from Eq.~\eqref{eq:angmom_conservation} to leading order in $\dot d$, we have
\begin{equation}
  \boldsymbol\tau^{\mathrm{matter}}
   =-\dot d\,\frac{\partial\mathbf L^{\mathrm{em}}_0}{\partial d}
    +\mathcal{O}(\dot d^2),
  \label{eq:tau_matter}
\end{equation}
where $\mathbf L^{\mathrm{em}}_0$ is the static field angular momentum expressed in Eq.~\eqref{eq:Lz_central}. The mechanical torque consists of the torque on the point charge and the torque on the sphere: $\boldsymbol\tau^{\mathrm{matter}}=\boldsymbol\tau^{q} +\boldsymbol\tau^{\mathrm{sphere}},$ where $\bm \tau^q=\int_{\mathbb{R}^3}d^3\bm r\,\bm r\times(\rho_\text{E} \bm E+{c}^{-1}\bm J_\text{E}\times\bm B)$, $\rho_\text{E}$ and $\bm J_\text{E}$ are the free sources appearing in Eq.~(\ref{eq:adiabatic}). One verifies that the former torque term vanishes to the order of interest: the point charge being on the symmetry axis, by azimuthal symmetry, the static electrostatic force on the charge is directed along $\hat{\mathbf z}$, and the correction of $\mathcal{O}(\dot d)$ induced by the motion either vanishes on the $z$-axis or remains directed along $\hat{\mathbf z}$, so that $\int_{\mathbb{R}^3}d^3\bm r\,\bm r\times(\rho_\text{E} \bm E_0+\rho_E\bm E_1+{c}^{-1}\bm J_\text{E}\times\bm B_0)=0$ and thus $\boldsymbol\tau^{q}=\mathcal{O}(\dot d^2)$. Hence, to leading order, $\boldsymbol{\tau}^{\mathrm{matter}}=\boldsymbol{\tau}^{\mathrm{sphere}}$, and
\begin{equation}
  \boldsymbol\tau^{\mathrm{sphere}}
   =-\dot d\,\frac{\partial\mathbf L^{\mathrm{em}}_0}{\partial d}
    +\mathcal{O}(\dot d^2),
  \label{eq:tau_sphere}
\end{equation}
which relates the torque acting on the sphere during the quasi-static motion of the charge along the $z$-axis to the derivative, with respect to $d$, of the static field angular momentum of Eq. (\ref{eq:Lz_central}). As in Feynman's disk paradox, a mechanical torque arises from an apparently static configuration as a direct consequence of the conservation of the total angular momentum. Using the fact that only the $z$ component of Eq. (\ref{eq:Lz_central}) is non-vanishing, we can rewrite Eq. (\ref{eq:tau_sphere}) as 
\begin{equation}\label{tau_z_sfera}
    \tau^{\mathrm{sphere}}_z=-\dot d\,\frac{\partial
L^{\mathrm{em}}_z}{\partial d}+\mathcal{O}(\dot{d}^2).
\end{equation}
We now treat the sphere as a rigid body with mass $M$ and moment of inertia $I=\tfrac{2}{5}MR^2$ \cite{goldstein_mechanics}, which is free to rotate about the $z$ axis, so that $\tau^{\mathrm{sphere}}_z=I\dot\omega$, where $\omega$ denotes its angular velocity. Integrating Eq. (\ref{tau_z_sfera}) over the whole process described by Eq.~\eqref{eq:d_of_t} and starting from rest, one obtains the angular velocity acquired at $t_1$:
\begin{equation}\label{omega(t_1)}
    \omega(t_1)=\frac{15}{8\pi R^5 D}\left[{L_z^\text{em}(d_0)-L_z^\text{em}(d_1)}\right],
\end{equation}
where we have written the mass in terms of the density $D$ of the sphere:
$M=\tfrac{4}{3}\pi R^3 D$, so that $I=\tfrac{8}{15}\pi R^5 D$. If $\Delta d/d_0 \ll 1$, one can write
\begin{equation}\label{eq_omega_approx}
    \omega(t_1)\approx-\frac{15}{8\pi R^5 D} {\frac{\partial L_z^\text{em}}{\partial d}\eval_{d=d_0}}{}\Delta d.
\end{equation}
The corresponding angular displacement follows from
$\vartheta(t_1)=\int_0^{t_1}\omega(t)\,dt$, which using Eq. (\ref{omega(t_1)}) gives the following
\begin{equation}\label{eq_theta(t_1)_esatta}
    \vartheta(t_1)=\frac{15}{8\pi R^5 D}\left[L_z^\text{em}(d_0)t_1-\int_0^{t_1}dt\,L_z^\text{em}(d(t))\right].
\end{equation}
If $\Delta d/d_0 \ll 1$, using Eq. (\ref{eq:d_of_t}), one can approximate $L_z^\text{em}(d(t))\approx L_z^\text{em}(d_0)+t\frac{\Delta d}{t_1}\partial_d L_z^\text{em}(d_0)$, obtaining
\begin{equation}
  \vartheta(t_1)\approx-\frac{15}{16\pi R^5 D} {\frac{\partial L_z^\text{em}}{\partial d}\eval_{d=d_0}}{}\Delta d\,t_1.
  \label{eq_theta(t_1)_approx}
\end{equation}
For fixed $R$, $D$, $d_0$, $d_1$ and $t_1$, Eqs.~\eqref{eq_omega_approx} and \eqref{eq_theta(t_1)_approx} imply that both $\omega(t_1)$ and $\vartheta(t_1)$ are substantially larger for a WSM than for a TI. The enhancement comes from two sources: the larger magnetoelectric factor, $\kappa_\text{WSM}=2\beta\sim10-100$ against $\kappa_\text{TI}=\Theta=\pi$, and the behavior of $|\partial_d\Lambda_{\mathrm{WSM}}|/\partial_d\Lambda_{\mathrm{TI}}$, which exceeds unity over essentially the entire range of $d/R$, as shown in Fig. \ref{fig:panels} (d).

 \begin{table}[t]
 \caption{\label{tab:estimates}Estimated angular velocity $\omega(t_1)$ and angular displacement $\vartheta(t_1)$ for the WSM and TI nanospheres, for the parameters given in the text.}
 \begin{ruledtabular}
 \begin{tabular}{cccc}
    & WSM & TI  \\ \colrule
   $\omega\,[\text{rad}/\text{s}]$ & $1\times10^{-3}$ & $-3\times10^{-6}$ \\
   $\vartheta\, [^{\circ}]$ & 5.8 & $-0.02$\\
 \end{tabular}
 \end{ruledtabular}
 \end{table}
Finally, we present a possible experimental realization. A WSM/TI nanosphere could be placed in the field of the charged tip of a colloidal-probe atomic force microscope (AFM) \cite{fleming_2013_afm_review,chighizola_2020_afm_colloidal_probes}, which acts, to a good approximation, as a point charge whose distance from the sphere can be varied quasi-statically with nanometer-level precision. Monitoring either the angular velocity or the angular displacement of the nanosphere could provide a direct signature of the electromagnetic angular-momentum transfer mediated by the axion magnetoelectric response. Using Eqs.~(\ref{eq:Lz_central}), (\ref{omega(t_1)}) and (\ref{eq_theta(t_1)_esatta}), with $N\sim10^4$ for the tip of a colloidal-probe AFM, for typical values $\varepsilon=10$, $R=150~\mathrm{nm}$, $2bR=75$, $d_0= 600~\mathrm{nm}$, $\Delta d= 10~\mu\mathrm{m}$, $D= 6\times 10^3~\mathrm{kg}/\mathrm{m}^3$, $t_1= 100~\mathrm{s}$, one finds the results shown in Table~\ref{tab:estimates}. If a scanning tunneling microscope (STM)~\cite{voigtlander_2018_stm_review} is employed, the effect is less visible since $N_\text{STM}\sim 10^2$.

Recent advances in levitated optomechanics have established these systems as promising platforms for investigating weak rotational phenomena \cite{hoang_2016_torsional_optomechanics,reimann_2018_ghz_rotation,ahn_2020_ultrasensitive_torque}. Their combination of excellent environmental isolation and high-sensitivity rotational readout makes them appealing candidates for probing the angular-momentum-transfer mechanism predicted here. However, a proper evaluation of experimental feasibility would require a dedicated study of rotational damping, optical restoring torques, electrostatic backgrounds, and additional sources of torque noise, which is beyond the scope of this work.

\section{Summary and Conclusions}\label{sec:summary_conclusions}
In this work, we have established a topological counterpart of Feynman's disk paradox by investigating the storage and transfer of electromagnetic angular momentum in a spherical Weyl semimetal interacting with a point charge.

Starting from the axion Maxwell's equations, we solved the electrostatic problem of a Weyl semimetal sphere to first order in the dimensionless coupling $\eta=2\alpha bR/\pi$, and compared the results with the exactly solvable case of a topological insulator sphere. While at leading order the electric response coincides with that of an ordinary dielectric sphere in both material classes, the magnetic response is markedly different. Unlike the topological insulator case, the spatially varying axion field in the Weyl semimetal makes the magnetic response depend on electric multipoles of different orders: the magnetic coefficients $A_n$ and $B_n$ couple to the electric coefficients $a_{n\pm1}$, while a non-harmonic contribution to the magnetic potential emerges from the effective bulk magnetoelectric sources generated by the spatially varying axion field.

From these solutions, we derived the electromagnetic angular momentum stored in the field, obtaining the general result of Eq.~(\ref{eq:Lz_central}). We found that the dimensionless factor $\Lambda$ exhibits qualitatively different behaviors in the two material classes: for the Weyl semimetal it contains only even powers of $R/d$ and decays as $(R/d)^4$, while for the topological insulator it contains only odd powers and decays as $(R/d)^5$. As a consequence, the Weyl semimetal exhibits a substantially larger electromagnetic angular momentum than the topological insulator. Moreover, our analysis revises the result of Ref.~\cite{feynman_paradox_ti}, showing that the leading asymptotic behavior for the topological insulator is $(R/d)^5$, rather than $(R/d)^3$, owing to a previously neglected contribution.

Finally, we showed that a quasi-static displacement of the external point charge converts the stored electromagnetic angular momentum into mechanical angular momentum, generating a mechanical torque on the sphere. The predicted angular velocity and, especially, the angular displacement are significantly higher for the Weyl semimetal than for the topological insulator, due to the larger topological factor $2bR$ and the stronger dependence of $\Lambda$ on the charge position $d/R$. More broadly, we have shown that axion electrodynamics not only modifies static electromagnetic responses but also gives rise to a mechanism for the storage and transfer of angular momentum. 


\section*{Acknowledgments}
The authors are grateful to F. Buccheri and E. Curcio for the discussions and insightful remarks provided during the initial phases of this work. A.B. expresses gratitude to G. Biondo and R. Ermito for their support. F.M.D.P. acknowledges the support of Università degli Studi di Catania through the project TCMQI PIACERI 2024/2026.

\appendix

\section{Legendre Polynomials}\label{sec:legendre_polynomials}

In this Appendix, we recall useful properties of the Legendre polynomials $P_n(x)$, which can be found in \cite{jackson_classical_electrodynamics,gradshteyn2007} or can be straightforwardly verified:

\begin{equation}\label{definizione_P_n(x)}
    \frac{d}{dx}\Big[(1-x^2)\frac{d}{dx}P_n(x)\Big]=-n(n+1)P_n(x),
\end{equation}
\begin{equation}\label{Legendre_orthogonality}
    \int_{-1}^1dx\,P_n(x)P_m(x)=\frac{2}{2n+1}\delta_{n,m},
\end{equation}
\begin{equation}\label{Legendre_espansione_1/r}
    \frac{1}{\sqrt{1-2xt+t^2}}=\sum_{n=0}^\infty t^nP_n(x), \, \, \, \, \, |x|\leq1,
\end{equation}
\begin{equation}\label{P_1(x)_P_n(x)}
    xP_n(x)=\frac{n+1}{2n+1}P_{n+1}(x)+\frac{n}{2n+1}P_{n-1}(x),
\end{equation}
\begin{equation}\label{derivata_P_n(x)}
    (1-x^2)P^\prime_n(x)=nP_{n-1}(x)-nxP_n(x),
\end{equation}
\begin{equation}\label{nabla^2_F_n(r)_P_n}
    \nabla^2[F_n(r)P_n(\cos\theta)]=[F_n^{\prime\prime}+\tfrac{2}{r}F_n^\prime-\tfrac{n(n+1)}{r^2}F_n]P_n(\cos\theta),
\end{equation}
\begin{equation}\label{partial_z_r^n_P_n}
    \frac{\partial}{\partial z}[r^nP_n(\cos\theta)]=nr^{n-1}P_{n-1}(\cos\theta).
\end{equation}
Let us now calculate the following useful integral, involving Legendre polynomials:
\begin{equation}
    I_{n,m}=\int_0^\pi d\theta\, \sin^3\theta\, P_n(\cos\theta)P^\prime_{m}(\cos\theta),
\end{equation}
where $P'_m(x)\equiv dP_m(x)/dx$.
It is convenient to introduce the variable $x=\cos\theta$, in order to write
\begin{equation}
    I_{n,m}=\int_{-1}^1 dx\,(1-x^2)P_n(x)P^\prime_{m}(x).
\end{equation}
Let us now apply Equations (\ref{P_1(x)_P_n(x)})-(\ref{derivata_P_n(x)}), getting
\begin{align}
I_{n,m}
 &=\int_{-1}^{1}\!dx\,P_n(x)\big[m P_{m-1}(x)-x\,m\,P_{m}(x)\big]\nonumber \\
 &=\int_{-1}^{1}\!dx\,P_n(x)\Big[m P_{m-1}(x)-\frac{m(m+1)}{2m+1}P_{m+1}(x)\nonumber \\
 &\hspace{3.8mm}\qquad\qquad\qquad-\frac{m^2}{2m+1}P_{m-1}(x)\Big].
\end{align}
After applying Equation (\ref{Legendre_orthogonality}) and performing some algebraic manipulations, we obtain the following result:
\begin{equation}\label{I_nm_Legendre}
    I_{n,m}=\frac{2}{2n+1}\left[\tfrac{(n+1)(n+2)}{2n+3}\delta_{m,n+1}-\tfrac{n(n-1)}{2n-1}\delta_{m,n-1}\right].
\end{equation}
Note that generally $I_{n,m}\neq I_{m,n}$. In particular, one has
\begin{equation}\label{I_mn_Legendre}
    I_{m,n}=\frac{2n(n+1)}{2n+1}\left(\frac{\delta_{m,n-1}}{2n-1}-\frac{\delta_{m,n+1}}{2n+3}\right).
\end{equation}

\vspace{2pt}
\section{Calculation of the electromagnetic angular momentum for a TI/WSM sphere and a point charge}\label{sec:calculation_em_appendix}

In this Appendix, we report the intermediate steps leading to Eqs.~\eqref{eq:Lz_central}, \eqref{eq:Lambda_WSM} and \eqref{eq:Lambda_TI} of Sec. \ref{sec:electromagnetic_angular_momentum_for_a_ti_wsm_sphere}. The calculation only relies on the general form of the potentials $\varphi$ and $\psi$, i.e.\ on the multipole expansions shown in Eqs.~\eqref{eq:WSM_phi_in} and \eqref{eq:WSM_psi_in}-\eqref{eq:WSM_psi_out}, without specifying the coefficients. It therefore applies to both material classes, TI and WSM. Starting from Eq.~\eqref{eq:Lz_integral} and separating the radial integration into the three regions $0<r<R$, $R<r<d$ and $r>d$, the trivial $\phi$ integration produces a factor $2\pi$ and we may write
\begin{equation}
  L_z^{\mathrm{em}}=\frac{1}{2c}
    \bigl(I_{1+}-I_{1-}+I_{2+}-I_{2-}+I_{3+}-I_{3-}\bigr),
  \label{eq:Lz_split}
\end{equation}
where we have defined
\begin{widetext}
\begin{subequations}\label{eq:I_all_def}
\begin{align}
  I_{1+}&=\int_0^R\!\!dr\int_0^\pi\!\!d\theta\;
    r^3\sin^2\!\theta\;E_{1r}H_{1\theta},\\[4pt]
    I_{1-}&=\int_0^R\!\!dr\int_0^\pi\!\!d\theta\;
    r^3\sin^2\!\theta\;E_{1\theta}H_{1r},\\[4pt]
  I_{2+}&=\int_R^d\!\!dr\int_0^\pi\!\!d\theta\;
    r^3\sin^2\!\theta\;E_{2r}H_{2\theta},\\[4pt]
    I_{2-}&=\int_R^d\!\!dr\int_0^\pi\!\!d\theta\;
    r^3\sin^2\!\theta\;E_{2\theta}H_{2r},\\[4pt]
  I_{3+}&=\int_d^{+\infty}\!\!dr\int_0^\pi\!\!d\theta\;
    r^3\sin^2\!\theta\;E_{2r}H_{2\theta},\\[4pt] I_{3-}&=\int_d^{+\infty}\!\!dr\int_0^\pi\!\!d\theta\;
    r^3\sin^2\!\theta\;E_{2\theta}H_{2r}.
\end{align}
\end{subequations}
Recalling that $\bm E=-\grad \varphi$ and $\bm H=-\grad \psi$, using the forms shown in Eqs. (\ref{eq:WSM_phi_in}) and (\ref{eq:WSM_psi_in})-(\ref{eq:WSM_psi_out}) of the scalar potentials $\varphi$ and $\psi$, and using the Legendre identities (\ref{I_nm_Legendre})-(\ref{I_mn_Legendre}) of Appendix~\ref{sec:legendre_polynomials} to perform the angular integrations, the six terms read
\begin{subequations}\label{eq:I_all_calcolati}
 
\begin{align}
  I_{1+}=-2\frac{\kappa}{\pi}\alpha q^{2}\sum_{n=0}^{\infty}\frac{n\,a_n}{2n+1}\Bigg[\,
    &\frac{(n+1)(n+2)}{2n+3}
      \bigg(A_{n+1}\frac{R^{2n+3}}{2n+3}+C_{n+1}\frac{R^{2n+5}}{2n+5}\bigg)\nonumber\\
      &-\frac{n(n-1)}{2n-1}
      \bigg(A_{n-1}\frac{R^{2n+1}}{2n+1}+C_{n-1}\frac{R^{2n+3}}{2n+3}\bigg)
      \Bigg],
  \label{eq:I1p}
\end{align}
 
\begin{align}
  I_{1-}=-2\frac{\kappa}{\pi}\alpha q^{2}\sum_{n=0}^{\infty}\frac{n(n+1)\,a_n}{2n+1}\Bigg\{\,
    &\frac{1}{2n-1}
      \bigg[(n-1)A_{n-1}\frac{R^{2n+1}}{2n+1}
        +(n+1)C_{n-1}\frac{R^{2n+3}}{2n+3}\bigg]\nonumber\\
    &-\frac{1}{2n+3}
      \bigg[(n+1)A_{n+1}\frac{R^{2n+3}}{2n+3}
        +(n+3)C_{n+1}\frac{R^{2n+5}}{2n+5}\bigg]\Bigg\},
  \label{eq:I1m}
\end{align}
 
\begin{align}
  I_{2+}=-2\frac{\kappa}{\pi} \alpha q^{2}\sum_{n=0}^{\infty}\frac{1}{2n+1}\Bigg\{\,
    &\frac{(n+1)(n+2)}{2n+3}B_{n+1}
      \bigg[\frac{n}{d^{n+1}}\ln\frac{d}{R}
      +(n+1)b_n\frac{d^{-(2n+1)}-R^{-(2n+1)}}{2n+1}\bigg]\nonumber\\
    &-\frac{n(n-1)}{2n-1}B_{n-1}
      \bigg[\frac{n}{d^{n+1}}\frac{d^{2}-R^{2}}{2}
      +(n+1)b_n\frac{d^{-(2n-1)}-R^{-(2n-1)}}{2n-1}\bigg]\Bigg\},
  \label{eq:I2p}
\end{align}
 
\begin{align}
  I_{2-}=2\frac{\kappa}{\pi} \alpha q^{2}\sum_{n=0}^{\infty}\frac{n(n+1)}{2n+1}\Bigg\{\,
    &\frac{n}{2n-1}B_{n-1}
      \bigg[\frac{1}{d^{n+1}}\frac{d^{2}-R^{2}}{2}
      -b_n\frac{d^{-(2n-1)}-R^{-(2n-1)}}{2n-1}\bigg]\nonumber\\
      &-\frac{n+2}{2n+3}B_{n+1}
      \bigg[\frac{1}{d^{n+1}}\ln\frac{d}{R}
     -b_n\frac{d^{-(2n+1)}-R^{-(2n+1)}}{2n+1}\bigg]\Bigg\},
  \label{eq:I2m}
\end{align}
 
\begin{align}
  I_{3+}=2\frac{\kappa}{\pi}\alpha q^{2}\sum_{n=0}^{\infty}\frac{(n+1)(d^{n}+b_n)}{2n+1}\bigg[\,
    &\frac{(n+1)(n+2)}{2n+3}B_{n+1}\frac{d^{-(2n+1)}}{2n+1}-\frac{n(n-1)}{2n-1}B_{n-1}\frac{d^{-(2n-1)}}{2n-1}\bigg],
  \label{eq:I3p}
\end{align}
 
\begin{align}
  I_{3-}=2\frac{\kappa}{\pi}\alpha q^{2}\sum_{n=0}^{\infty}\frac{n(n+1)(d^{n}+b_n)}{2n+1}\bigg[\,
    &\frac{n}{2n-1}B_{n-1}\frac{d^{-(2n-1)}}{2n-1}-\frac{n+2}{2n+3}B_{n+1}\frac{d^{-(2n+1)}}{2n+1}\bigg],
  \label{eq:I3m}
\end{align}
\end{subequations}
where we have used $\kappa=2\beta=2bR$ for the WSM and $\kappa=\Theta=\pi$ for the TI. Let us define
\begin{equation}
    \Lambda_i=({I_{i+}-I_{i-}})/({2 \frac{\kappa}{\pi}\alpha q^2}),\qquad i=1,2,3,
\end{equation}
and $\Lambda=\Lambda_1+\Lambda_2+\Lambda_3$, so that
$L_z^{\mathrm{em}}=\frac{\kappa}{\pi}q^2\frac{\alpha}{c}\Lambda$, which reproduces
Eq.~\eqref{eq:Lz_central} upon using $q=Ne$ and
$e^2/c=\alpha\hbar$. Reorganizing the dummy indices of the sums
in Eqs.~\eqref{eq:I1p}-\eqref{eq:I3m}, one finds the general expressions for the three $\Lambda_i$:
 
\begin{align}
  \Lambda_1&=\sum_{n=0}^{\infty}\frac{n(n+1)}{(2n+1)(2n+3)}
    \left[(A_n a_{n+1}-A_{n+1}a_n)\,R^{2n+3}
     +\left(\frac{2n^2+5n+4}{2n^2+5n}\,C_n a_{n+1}-C_{n+1}a_n\right)R^{2n+5}
    \right],
  \label{eq:Lambda1_gen}\\[6pt]
  \Lambda_2&=\sum_{n=0}^{\infty}
    \left\{\frac{(n+1)(n+2)}{(2n+1)(2n+3)}
     \frac{B_{n+1}b_n-B_n b_{n+1}}{R^{2n+1}}
     \left[1-\left(\frac{R}{d}\right)^{2n+1}\right]
     -\frac{n^2}{(2n-1)(2n+1)}\frac{B_{n-1}}{d^{n-1}}
     \left[1-\left(\frac{R}{d}\right)^{2}\right]\right\},
  \label{eq:Lambda2_gen}\\[6pt]
  \Lambda_3&=\sum_{n=0}^{\infty}
    \left(1+\frac{b_n}{d^n}\right)
    \left[\frac{(n+1)(n+2)}{(2n+1)(2n+3)}\frac{B_{n+1}}{d^{n+1}}
     -\frac{n(n+1)}{(2n-1)(2n+1)}\frac{B_{n-1}}{d^{n-1}}\right].
  \label{eq:Lambda3_gen}
\end{align}
We emphasize that Eqs.~\eqref{eq:Lambda1_gen}-\eqref{eq:Lambda3_gen} are
general: they only assume the multipole forms shown in Eqs.~(\ref{eq:WSM_phi_in}) and (\ref{eq:WSM_psi_in})-(\ref{eq:WSM_psi_out}) of $\varphi$ and $\psi$, and hold for both the WSM and the TI once the corresponding coefficients are inserted. 
Focusing on Eqs.~(\ref{eq:Lambda1_gen})-(\ref{eq:Lambda3_gen}) for the WSM case, upon substituting into them the WSM coefficients~\eqref{eq:aWSM}-\eqref{eq:CWSM} and reorganizing the terms, one finds
\begin{align}
  &\Lambda_{1,\mathrm{WSM}}
    =-\sum_{n=0}^{\infty}
       \frac{2(n+1)^2(2n+3)}{(2n+1)^2(2n+5)^2\,\varepsilon_{n+1}^2}
       \left(\frac{R}{d}\right)^{2n+4}+\mathcal{O}(\eta^2),
  \label{eq:Lambda1_WSM}\\
  &\Lambda_{2,\mathrm{WSM}}+\Lambda_{3,\mathrm{WSM}}
    =\varepsilon\sum_{n=0}^{\infty}
       \frac{2(n+1)^2(2n+3)}{(2n+1)^2(2n+5)^2\,\varepsilon_{n+1}^2}
       \left(\frac{R}{d}\right)^{2n+4}+\mathcal{O}(\eta^2).
  \label{eq:Lambda23_WSM}
\end{align}
Summing Eqs.~(\ref{eq:Lambda1_WSM})-(\ref{eq:Lambda23_WSM}) yields $\Lambda_{\mathrm{WSM}}$ of Eq. \eqref{eq:Lambda_WSM}.

For the TI, we insert into Eqs.~\eqref{eq:Lambda1_gen}-\eqref{eq:Lambda3_gen} the leading order coefficients reported in Eqs.~ (\ref{a_n_TI_1storder})-(\ref{A_n_B_n_TI_1storder}) together with Eq. (\ref{C_n_TI_esatto}). Since $b_n^{\mathrm{TI}}$ and $B_n^{\mathrm{TI}}$ differ only by the factor $(\varepsilon-1)$ [cf.\
Eqs.~(\ref{b_n_TI_1storder})-(\ref{A_n_B_n_TI_1storder})], the combination
$B_{n+1}b_n-B_n b_{n+1}$ vanishes and the first term of
Eq.~\eqref{eq:Lambda2_gen} drops out, exactly recovering Equation (54) of Ref. \cite{feynman_paradox_ti}:
\begin{align}
    \Lambda_\text{2,TI}+\Lambda_\text{3,TI}=\sum_{n=0}^\infty \Big\{&-\frac{n^2}{(2n-1)(2n+1)}\frac{B^{\mathrm{TI}}_{n-1}}{d^{n-1}}\Big[1-\Big(\frac{R}{d}\Big)^{2}\Big]\notag\\
    &+\Big(1+\frac{b^{\mathrm{TI}}_n}{d^n}\Big)\Big[\frac{(n+1)(n+2)}{(2n+1)(2n+3)}\frac{B^{\mathrm{TI}}_{n+1}}{d^{n+1}}-\frac{n(n+1)}{(2n-1)(2n+1)}\frac{B^{\mathrm{TI}}_{n-1}}{d^{n-1}}\Big]\Big\}+\mathcal{O}(\alpha^2)\label{Lambda_TI_riportata_nel_paper}.
\end{align}
When performing this sum, one may check that the contributions proportional to $(R/d)^3$ cancel among themselves: this is the origin of the discrepancy between Eq. (\ref{eq:Lambda_TI_asy}) and Equation (57) of Ref.~\cite{feynman_paradox_ti}. Moreover, the internal contribution left out in Ref.~\cite{feynman_paradox_ti} does not vanish, since $A_n^{\mathrm{TI}}a_{n+1}^{\mathrm{TI}}-A_{n+1}^{\mathrm{TI}}a_n^{\mathrm{TI}}\neq0$. Carrying out the calculations and rearranging the terms, one obtains
\begin{align}
&\Lambda_{1,\mathrm{TI}}
    =\sum_{n=0}^{\infty}
     \frac{(n+1)(n+2)}{(2n+3)(2n+5)\,\varepsilon_{n+1}\varepsilon_{n+2}}
     \left(\frac{R}{d}\right)^{2n+5}+\mathcal{O}(\alpha^2),
  \label{eq:Lambda1_TI}\\
    &\Lambda_\text{2,TI}+\Lambda_\text{3,TI}=-\varepsilon\sum_{n=0}^\infty \frac{(n+1)(n+2)}{(2n+3)(2n+5)\varepsilon_{n+1}\varepsilon_{n+2}}\,\Big(\frac{R}{d}\Big)^{2n+5}+\mathcal{O}(\alpha^2).\label{eq:Lambda23_TI}
\end{align}
Summing Eqs.~\eqref{eq:Lambda1_TI}-\eqref{eq:Lambda23_TI} gives
$\Lambda_{\mathrm{TI}}$ of Eq.~\eqref{eq:Lambda_TI}, we notice that without the contribution $\Lambda_\text{TI}$ inside the sphere would not vanish for $\varepsilon=1$. We also report the derivatives of $\Lambda$ with respect to $d$ in the two cases:
\begin{align}
  \frac{\partial\Lambda_{\mathrm{WSM}}}{\partial d}
    &=-\frac{\varepsilon-1}{d}\sum_{n=0}^{\infty}
     \frac{4(n+1)^2(n+2)(2n+3)}{(2n+1)^2(2n+5)^2\,\varepsilon_{n+1}^2}
     \left(\frac{R}{d}\right)^{2n+4}+\mathcal{O}(\eta^2),
  \label{eq:dLambda_WSM}\\
  \frac{\partial\Lambda_{\mathrm{TI}}}{\partial d}
    &=\frac{\varepsilon-1}{d}\sum_{n=0}^{\infty}
     \frac{(n+1)(n+2)}{(2n+3)\,\varepsilon_{n+1}\varepsilon_{n+2}}
     \left(\frac{R}{d}\right)^{2n+5}+\mathcal{O}(\alpha^2).
  \label{eq:dLambda_TI}
\end{align}
\end{widetext}

\bibliography{biblio}

@article{weyl_1929_electron_and_gravitation,
    author = "Weyl, H.",
    title = "{Electron and Gravitation. 1. (In German)}",
    doi = "10.1007/BF01339504",
    journal = "Z. Phys.",
    volume = "56",
    pages = "330--352",
    year = "1929"}

@article{nielsen_ninomiya_1981_proof1,
title = {Absence of neutrinos on a lattice: (I). Proof by homotopy theory},
journal = {Nuclear Physics B},
volume = {185},
number = {1},
pages = {20-40},
year = {1981},
issn = {0550-3213},
doi = {https://doi.org/10.1016/0550-3213(81)90361-8},
url = {https://www.sciencedirect.com/science/article/pii/0550321381903618},
author = {H.B. Nielsen and M. Ninomiya},}

@article{nielsen_ninomiya_1981_proof2,
title = {Absence of neutrinos on a lattice: (II). Intuitive topological proof},
journal = {Nuclear Physics B},
volume = {193},
number = {1},
pages = {173-194},
year = {1981},
issn = {0550-3213},
doi = {https://doi.org/10.1016/0550-3213(81)90524-1},
url = {https://www.sciencedirect.com/science/article/pii/0550321381905241},
author = {H.B. Nielsen and M. Ninomiya},}

@article{berry_1983_geometric_phase,
    author = {Berry, Michael Victor},
    title = {Quantal phase factors accompanying adiabatic changes},
    journal = {Proceedings of the Royal Society of London. A. Mathematical and Physical Sciences},
    volume = {392},
    number = {1802},
    pages = {45-57},
    year = {1984},
    month = {03},
    issn = {0080-4630},
    doi = {10.1098/rspa.1984.0023},
    url = {https://doi.org/10.1098/rspa.1984.0023},
}

@article{xiao_2010_berry_phase_review,
  title = {Berry phase effects on electronic properties},
  author = {Xiao, Di and Chang, Ming-Che and Niu, Qian},
  journal = {Rev. Mod. Phys.},
  volume = {82},
  issue = {3},
  pages = {1959--2007},
  numpages = {0},
  year = {2010},
  month = {Jul},
  publisher = {American Physical Society},
  doi = {10.1103/RevModPhys.82.1959},
  url = {https://link.aps.org/doi/10.1103/RevModPhys.82.1959}}

@article{wan_2011_topological_semimetals_and_fermi_arc_surface_states,
  title = {Topological semimetal and Fermi-arc surface states in the electronic structure of pyrochlore iridates},
  author = {Wan, Xiangang and Turner, Ari M. and Vishwanath, Ashvin and Savrasov, Sergey Y.},
  journal = {Phys. Rev. B},
  volume = {83},
  issue = {20},
  pages = {205101},
  numpages = {9},
  year = {2011},
  month = {May},
  publisher = {American Physical Society},
  doi = {10.1103/PhysRevB.83.205101},
  url = {https://link.aps.org/doi/10.1103/PhysRevB.83.205101}}

@article{yang_2011_qhe_in_wsm,
  title = {Quantum Hall effects in a Weyl semimetal: Possible application in pyrochlore iridates},
  author = {Yang, Kai-Yu and Lu, Yuan-Ming and Ran, Ying},
  journal = {Phys. Rev. B},
  volume = {84},
  issue = {7},
  pages = {075129},
  numpages = {11},
  year = {2011},
  month = {Aug},
  publisher = {American Physical Society},
  doi = {10.1103/PhysRevB.84.075129},
  url = {https://link.aps.org/doi/10.1103/PhysRevB.84.075129}}

@article{burkov_2011_wsm_in_ti_layers,
  title = {Weyl Semimetal in a Topological Insulator Multilayer},
  author = {Burkov, A. A. and Balents, Leon},
  journal = {Phys. Rev. Lett.},
  volume = {107},
  issue = {12},
  pages = {127205},
  numpages = {4},
  year = {2011},
  month = {Sep},
  publisher = {American Physical Society},
  doi = {10.1103/PhysRevLett.107.127205},
  url = {https://link.aps.org/doi/10.1103/PhysRevLett.107.127205}}

@article{burkov_2011_topological_nodal_semimetals,
  title = {Topological nodal semimetals},
  author = {Burkov, A. A. and Hook, M. D. and Balents, Leon},
  journal = {Phys. Rev. B},
  volume = {84},
  issue = {23},
  pages = {235126},
  numpages = {14},
  year = {2011},
  month = {Dec},
  publisher = {American Physical Society},
  doi = {10.1103/PhysRevB.84.235126},
  url = {https://link.aps.org/doi/10.1103/PhysRevB.84.235126}}

@article{lv_2015_observation_nodes_taas,
  author  = {Lv, B. Q. and Xu, N. and Weng, H. M. and Ma, J. Z. and Richard, P. and Huang, X. C. and Zhao, L. X. and Chen, G. F. and Matt, C. E. and Bisti, F. and Strocov, V. N. and Mesot, J. and Fang, Z. and Dai, X. and Qian, T. and Shi, M. and Ding, H.},
  title   = {Observation of Weyl nodes in TaAs},
  journal = {Nature Physics},
  year    = {2015},
  volume  = {11},
  number  = {9},
  pages   = {724--727},
  doi     = {10.1038/nphys3426},
  url     = {https://doi.org/10.1038/nphys3426}}

@article{
xu_2015_observation_of_fermi_arcs_in_na3_bi,
author = {Su-Yang Xu  and Chang Liu  and Satya K. Kushwaha  and Raman Sankar  and Jason W. Krizan  and Ilya Belopolski  and Madhab Neupane  and Guang Bian  and Nasser Alidoust  and Tay-Rong Chang  and Horng-Tay Jeng  and Cheng-Yi Huang  and Wei-Feng Tsai  and Hsin Lin  and Pavel P. Shibayev  and Fang-Cheng Chou  and Robert J. Cava  and M. Zahid Hasan },
title = {Observation of Fermi arc surface states in a topological metal},
journal = {Science},
volume = {347},
number = {6219},
pages = {294-298},
year = {2015},
doi = {10.1126/science.1256742},
URL = {https://www.science.org/doi/abs/10.1126/science.1256742},
}

@article{huang_2015_negative_magnetoresistance_taas,
  title = {Observation of the Chiral-Anomaly-Induced Negative Magnetoresistance in 3D Weyl Semimetal TaAs},
  author = {Huang, Xiaochun and Zhao, Lingxiao and Long, Yujia and Wang, Peipei and Chen, Dong and Yang, Zhanhai and Liang, Hui and Xue, Mianqi and Weng, Hongming and Fang, Zhong and Dai, Xi and Chen, Genfu},
  journal = {Phys. Rev. X},
  volume = {5},
  issue = {3},
  pages = {031023},
  numpages = {9},
  year = {2015},
  month = {Aug},
  publisher = {American Physical Society},
  doi = {10.1103/PhysRevX.5.031023},
  url = {https://link.aps.org/doi/10.1103/PhysRevX.5.031023}}

@article{pellegrino_2015_helicons_wsm,
  title = {Helicons in Weyl semimetals},
  author = {Pellegrino, Francesco M. D. and Katsnelson, Mikhail I. and Polini, Marco},
  journal = {Phys. Rev. B},
  volume = {92},
  issue = {20},
  pages = {201407(R)},
  numpages = {5},
  year = {2015},
  month = {Nov},
  publisher = {American Physical Society},
  doi = {10.1103/PhysRevB.92.201407},
  url = {https://link.aps.org/doi/10.1103/PhysRevB.92.201407}}

@article{yan_felser_2017_review_wsm,
   author = "Yan, Binghai and Felser, Claudia",
   title = "Topological Materials: Weyl Semimetals", 
   journal= "Annual Review of Condensed Matter Physics",
   year = "2017",
   volume = "8",
   number = "Volume 8, 2017",
   pages = "337-354",
   doi = "https://doi.org/10.1146/annurev-conmatphys-031016-025458",
   url = "https://www.annualreviews.org/content/journals/10.1146/annurev-conmatphys-031016-025458",
   publisher = "Annual Reviews",
   issn = "1947-5462",
   type = "Journal Article",}

@article{hasan_xu_2017_review_wsm,
   author = "Hasan, M. Zahid and Xu, Su-Yang and Belopolski, Ilya and Huang, Shin-Ming",
   title = "Discovery of Weyl Fermion Semimetals and Topological Fermi Arc States", 
   journal= "Annual Review of Condensed Matter Physics",
   year = "2017",
   volume = "8",
   number = "Volume 8, 2017",
   pages = "289-309",
   doi = "https://doi.org/10.1146/annurev-conmatphys-031016-025225",
   url = "https://www.annualreviews.org/content/journals/10.1146/annurev-conmatphys-031016-025225",
   publisher = "Annual Reviews",
   issn = "1947-5462",
   type = "Journal Article",}

@article{armitage_mele_2018_review_wsm,
  title = {Weyl and Dirac semimetals in three-dimensional solids},
  author = {Armitage, N. P. and Mele, E. J. and Vishwanath, Ashvin},
  journal = {Rev. Mod. Phys.},
  volume = {90},
  issue = {1},
  pages = {015001},
  numpages = {57},
  year = {2018},
  month = {Jan},
  publisher = {American Physical Society},
  doi = {10.1103/RevModPhys.90.015001},
  url = {https://link.aps.org/doi/10.1103/RevModPhys.90.015001}}

@ARTICLE{xu_liu_2018_fermiarcs_in_co3sn2,
       author = {{Xu}, Qiunan and {Liu}, Enke and {Shi}, Wujun and {Muechler}, Lukas and {Gayles}, Jacob and {Felser}, Claudia and {Sun}, Yan},
        title = "{Topological surface Fermi arcs in the magnetic Weyl semimetal Co$_{3}$Sn$_{2}$S$_{2}$}",
      journal = {\prb},
         year = 2018,
        month = jun,
       volume = {97},
       number = {23},
          eid = {235416},
        pages = {235416},
          doi = {10.1103/PhysRevB.97.235416},
       adsurl = {https://ui.adsabs.harvard.edu/abs/2018PhRvB..97w5416X}
}

@article{xu_2020_electronic_correlations_in_co3sn2s2,
author = {Xu, Yueshan and Zhao, Jianzhou and Yi, Changjiang and Wang, Qi and Yin, Qiangwei and Wang, Yilin and Hu, Xiaolei and Wang, Luyang and Liu, Enke and Xu, Gang and Lu, Ling and Soluyanov, Alexey and Lei, Hechang and Shi, Youguo and Luo, Jianlin and Chen, Zhi-Guo},
year = {2020},
month = {08},
pages = {3985},
title = {Electronic correlations and flattened band in magnetic Weyl semimetal candidate ${\mathrm{Co}}_{3}{\mathrm{Sn}}_{2}{\mathrm{S}}_{2}$},
volume = {11},
journal = {Nature Communications},
doi = {10.1038/s41467-020-17234-0}}

@article{zu_2021_optical_properties_taas_nbas,
  title = {Comprehensive anisotropic linear optical properties of the Weyl semimetals TaAs and NbAs},
  author = {Zu, Rui and Gu, Mingqiang and Min, Lujin and Hu, Chaowei and Ni, Ni and Mao, Zhiqiang and Rondinelli, James M. and Gopalan, Venkatraman},
  journal = {Phys. Rev. B},
  volume = {103},
  issue = {16},
  pages = {165137},
  numpages = {17},
  year = {2021},
  month = {Apr},
  publisher = {American Physical Society},
  doi = {10.1103/PhysRevB.103.165137},
  url = {https://link.aps.org/doi/10.1103/PhysRevB.103.165137}}

@article{martin_ruiz_2019_charge_near_wsm,
  title = {Electromagnetic fields induced by an electric charge near a Weyl semimetal},
  author = {Mart\'{\i}n-Ruiz, A. and Cambiaso, M. and Urrutia, L. F.},
  journal = {Phys. Rev. B},
  volume = {99},
  issue = {15},
  pages = {155142},
  numpages = {13},
  year = {2019},
  month = {Apr},
  publisher = {American Physical Society},
  doi = {10.1103/PhysRevB.99.155142},
  url = {https://link.aps.org/doi/10.1103/PhysRevB.99.155142}
}

@article{bonasera_pellegrino_2022_tunable_floquet_wsm,
  title = {Tunable interface states between Floquet-Weyl semimetals},
  author = {Bonasera, F. and Zhang, S.-B. and Privitera, L. and Pellegrino, F. M. D.},
  journal = {Phys. Rev. B},
  volume = {106},
  issue = {19},
  pages = {195115},
  numpages = {13},
  year = {2022},
  month = {Nov},
  publisher = {American Physical Society},
  doi = {10.1103/PhysRevB.106.195115},
  url = {https://link.aps.org/doi/10.1103/PhysRevB.106.195115}
}

@article{wilczek_applications_axion_electrodynamics,
  title = {Two applications of axion electrodynamics},
  author = {Wilczek, Frank},
  journal = {Phys. Rev. Lett.},
  volume = {58},
  issue = {18},
  pages = {1799--1802},
  numpages = {0},
  year = {1987},
  month = {May},
  publisher = {American Physical Society},
  doi = {10.1103/PhysRevLett.58.1799},
  url = {https://link.aps.org/doi/10.1103/PhysRevLett.58.1799}}

@article{zyuzin_burkov_2012_wsm_response,
  title = {Topological response in Weyl semimetals and the chiral anomaly},
  author = {Zyuzin, A. A. and Burkov, A. A.},
  journal = {Phys. Rev. B},
  volume = {86},
  issue = {11},
  pages = {115133},
  numpages = {8},
  year = {2012},
  month = {Sep},
  publisher = {American Physical Society},
  doi = {10.1103/PhysRevB.86.115133},
  url = {https://link.aps.org/doi/10.1103/PhysRevB.86.115133}}

@article{vazifeh_2013_em_response_wsm,
  title = {Electromagnetic Response of Weyl Semimetals},
  author = {Vazifeh, M. M. and Franz, M.},
  journal = {Phys. Rev. Lett.},
  volume = {111},
  issue = {2},
  pages = {027201},
  numpages = {5},
  year = {2013},
  month = {Jul},
  publisher = {American Physical Society},
  doi = {10.1103/PhysRevLett.111.027201},
  url = {https://link.aps.org/doi/10.1103/PhysRevLett.111.027201}}

@article{goswami_2013_3+1_field_teory_wsm,
  title = {Axionic field theory of $(3+1)$-dimensional Weyl semimetals},
  author = {Goswami, Pallab and Tewari, Sumanta},
  journal = {Phys. Rev. B},
  volume = {88},
  issue = {24},
  pages = {245107},
  numpages = {9},
  year = {2013},
  month = {Dec},
  publisher = {American Physical Society},
  doi = {10.1103/PhysRevB.88.245107},
  url = {https://link.aps.org/doi/10.1103/PhysRevB.88.245107}}

@article{zyuzin_2015_chiral_em_wave_in_wsm,
  title = {Chiral electromagnetic waves in Weyl semimetals},
  author = {Zyuzin, Alexander A. and Zyuzin, Vladimir A.},
  journal = {Phys. Rev. B},
  volume = {92},
  issue = {11},
  pages = {115310},
  numpages = {4},
  year = {2015},
  month = {Sep},
  publisher = {American Physical Society},
  doi = {10.1103/PhysRevB.92.115310},
  url = {https://link.aps.org/doi/10.1103/PhysRevB.92.115310}}

@article{kotov_2018_tunable_nonreciprocity_of_light_in_wsm,
  title = {Giant tunable nonreciprocity of light in Weyl semimetals},
  author = {Kotov, O. V. and Lozovik, Yu. E.},
  journal = {Phys. Rev. B},
  volume = {98},
  issue = {19},
  pages = {195446},
  numpages = {9},
  year = {2018},
  month = {Nov},
  publisher = {American Physical Society},
  doi = {10.1103/PhysRevB.98.195446},
  url = {https://link.aps.org/doi/10.1103/PhysRevB.98.195446}}

@article{sekine_axion_2021_ed_in_topological_materials,
    author = {Sekine, Akihiko and Nomura, Kentaro},
    title = {Axion electrodynamics in topological materials},
    journal = {Journal of Applied Physics},
    volume = {129},
    number = {14},
    pages = {141101},
    year = {2021},
    month = {04},
    issn = {0021-8979},
    doi = {10.1063/5.0038804},
    url = {https://doi.org/10.1063/5.0038804},
    }

@article{zhou_2015_plasmon_detecting_chirality,
  title = {Plasmon mode as a detection of the chiral anomaly in Weyl semimetals},
  author = {Zhou, Jianhui and Chang, Hao-Ran and Xiao, Di},
  journal = {Phys. Rev. B},
  volume = {91},
  issue = {3},
  pages = {035114},
  numpages = {8},
  year = {2015},
  month = {Jan},
  publisher = {American Physical Society},
  doi = {10.1103/PhysRevB.91.035114},
  url = {https://link.aps.org/doi/10.1103/PhysRevB.91.035114}}

@article{song_2017_fermi_arc_plasmons_wsm,
  title = {Fermi arc plasmons in Weyl semimetals},
  author = {Song, Justin C. W. and Rudner, Mark S.},
  journal = {Phys. Rev. B},
  volume = {96},
  issue = {20},
  pages = {205443},
  numpages = {8},
  year = {2017},
  month = {Nov},
  publisher = {American Physical Society},
  doi = {10.1103/PhysRevB.96.205443},
  url = {https://link.aps.org/doi/10.1103/PhysRevB.96.205443}}

@article{andolina_pellegrino_2018_nonlocal_theory_plasmons_wsm,
  title = {Quantum nonlocal theory of topological Fermi arc plasmons in Weyl semimetals},
  author = {Andolina, Gian Marcello and Pellegrino, Francesco M. D. and Koppens, Frank H. L. and Polini, Marco},
  journal = {Phys. Rev. B},
  volume = {97},
  issue = {12},
  pages = {125431},
  numpages = {12},
  year = {2018},
  month = {Mar},
  publisher = {American Physical Society},
  doi = {10.1103/PhysRevB.97.125431},
  url = {https://link.aps.org/doi/10.1103/PhysRevB.97.125431}}

@article{tamaya_2019_surface_plasmons_film_wsm,
doi = {10.1088/1361-648X/ab17b3},
url = {https://doi.org/10.1088/1361-648X/ab17b3},
year = {2019},
month = {may},
publisher = {IOP Publishing},
volume = {31},
number = {30},
pages = {305001},
author = {Tamaya, Tomohiro and Kato, Takeo and Tsuchikawa, Kota and Konabe, Satoru and Kawabata, Shiro},
title = {Surface plasmon polaritons in thin-film Weyl semimetals},
journal = {Journal of Physics: Condensed Matter},}

@article{tsuchikawa_2020_wsm_plasmon_polaritons,
  title = {Characterization of a Weyl semimetal using a unique feature of surface plasmon polaritons},
  author = {Tsuchikawa, Kota and Konabe, Satoru and Yamamoto, Takahiro and Kawabata, Shiro},
  journal = {Phys. Rev. B},
  volume = {102},
  issue = {3},
  pages = {035443},
  numpages = {8},
  year = {2020},
  month = {Jul},
  publisher = {American Physical Society},
  doi = {10.1103/PhysRevB.102.035443},
  url = {https://link.aps.org/doi/10.1103/PhysRevB.102.035443}}

@article{bugaiko_2020_surface_plasmons_strained_wsm,
  title = {Surface plasmon polaritons in strained Weyl semimetals},
  author = {Bugaiko, O. V. and Gorbar, E. V. and Sukhachov, P. O.},
  journal = {Phys. Rev. B},
  volume = {102},
  issue = {8},
  pages = {085426},
  numpages = {13},
  year = {2020},
  month = {Aug},
  publisher = {American Physical Society},
  doi = {10.1103/PhysRevB.102.085426},
  url = {https://link.aps.org/doi/10.1103/PhysRevB.102.085426}}

@article{heidari_2021_anomalous_plasmons_strained_wsm,
  title = {Anomalous plasmon mode in strained Weyl semimetals},
  author = {Heidari, Shiva and Culcer, Dimitrie and Asgari, Reza},
  journal = {Phys. Rev. B},
  volume = {103},
  issue = {3},
  pages = {035306},
  numpages = {11},
  year = {2021},
  month = {Jan},
  publisher = {American Physical Society},
  doi = {10.1103/PhysRevB.103.035306},
  url = {https://link.aps.org/doi/10.1103/PhysRevB.103.035306}}

@article{peluso_buccheri_2025_wsm_waveguides,
  title = {Nonreciprocal Weyl semimetal waveguide},
  author = {Peluso, Marco and De Martino, Alessandro and Egger, Reinhold and Buccheri, Francesco},
  journal = {Phys. Rev. Res.},
  volume = {7},
  issue = {2},
  pages = {023195},
  numpages = {11},
  year = {2025},
  month = {May},
  publisher = {American Physical Society},
  doi = {10.1103/PhysRevResearch.7.023195},
  url = {https://link.aps.org/doi/10.1103/PhysRevResearch.7.023195}}

@article{pellegrino_2025_surface_plasmons_wsm,
  title = {Localized surface plasmons in a Weyl semimetal nanosphere},
  author = {Pellegrino, Francesco M. D. and Buccheri, Francesco and Angilella, G. G. N.},
  journal = {Phys. Rev. B},
  volume = {112},
  issue = {7},
  pages = {075431},
  numpages = {12},
  year = {2025},
  month = {Aug},
  publisher = {American Physical Society},
  doi = {10.1103/7b1l-gkjy},
  url = {https://link.aps.org/doi/10.1103/7b1l-gkjy}}

@article{xu_2011_wsm_hgcr2se4,
  title = {Chern Semimetal and the Quantized Anomalous Hall Effect in ${\mathrm{HgCr}}_{2}{\mathrm{Se}}_{4}$},
  author = {Xu, Gang and Weng, Hongming and Wang, Zhijun and Dai, Xi and Fang, Zhong},
  journal = {Phys. Rev. Lett.},
  volume = {107},
  issue = {18},
  pages = {186806},
  numpages = {5},
  year = {2011},
  month = {Oct},
  publisher = {American Physical Society},
  doi = {10.1103/PhysRevLett.107.186806},
  url = {https://link.aps.org/doi/10.1103/PhysRevLett.107.186806}}

@article{gao_2023_wsm_mnx2b2t6,
  title = {Intrinsic ferromagnetic axion states and single pair of Weyl fermions in the stable-state $\mathrm{MnX}_{2}\mathrm{B}_{2}\mathrm{T}_{6}$ family of materials},
  author = {Gao, Yan and Wu, Weikang and Gong, Ben-Chao and Yang, Huan-Cheng and Zhou, Xiang-Feng and Liu, Yong and Yang, Shengyuan A. and Liu, Kai and Lu, Zhong-Yi},
  journal = {Phys. Rev. B},
  volume = {107},
  issue = {4},
  pages = {045136},
  numpages = {7},
  year = {2023},
  month = {Jan},
  publisher = {American Physical Society},
  doi = {10.1103/PhysRevB.107.045136},
  url = {https://link.aps.org/doi/10.1103/PhysRevB.107.045136}}

@article{boulton_2024_searching_antiferromagnetic_wsm,
doi = {10.1088/1361-648X/ad5d3c},
url = {https://doi.org/10.1088/1361-648X/ad5d3c},
year = {2024},
month = {jul},
publisher = {IOP Publishing},
volume = {36},
number = {40},
pages = {405601},
author = {Boulton, James A and Kim, Ki Wook},
title = {Search for an antiferromagnetic Weyl semimetal in $\mathrm{(MnTe)}_{m}\mathrm{(\mathrm{Sb}_2\mathrm{Te}_3)}_{n}$ and $\mathrm{(MnTe)}_{m}(\mathrm{Bi}_2\mathrm{Te}_3)_{n}$ superlattices},
journal = {Journal of Physics: Condensed Matter},}

@article{liu_2024_wsm_kcrte_rbcrte,
  title = {Ideal spin-polarized Weyl half-semimetal with a single pair of Weyl points in the half-Heusler compounds $X\mathrm{CrTe}$ ($X=\mathrm{K}$, Rb)},
  author = {Liu, Hongshuang and Cao, Jin and Zhang, Zeying and Liang, Jiashuo and Wang, Liying and Yang, Shengyuan A.},
  journal = {Phys. Rev. B},
  volume = {109},
  issue = {17},
  pages = {174426},
  numpages = {7},
  year = {2024},
  month = {May},
  publisher = {American Physical Society},
  doi = {10.1103/PhysRevB.109.174426},
  url = {https://link.aps.org/doi/10.1103/PhysRevB.109.174426}}

@article{sushkov_2015_wsm_eu2ir2o7,
  title = {Optical evidence for a Weyl semimetal state in pyrochlore ${\mathrm{Eu}}_{2}{\mathrm{Ir}}_{2}{\mathrm{O}}_{7}$},
  author = {Sushkov, A. B. and Hofmann, J. B. and Jenkins, G. S. and Ishikawa, J. and Nakatsuji, S. and Das Sarma, S. and Drew, H. D.},
  journal = {Phys. Rev. B},
  volume = {92},
  issue = {24},
  pages = {241108(R)},
  numpages = {4},
  year = {2015},
  month = {Dec},
  publisher = {American Physical Society},
  doi = {10.1103/PhysRevB.92.241108},
  url = {https://link.aps.org/doi/10.1103/PhysRevB.92.241108}}

@article{cao_2023_wsm_mn3sn,
  title = {Optical study of the three-dimensional Weyl semimetal ${\mathrm{Mn}}_{3}\mathrm{Sn}$},
  author = {Cao, L. Y. and Xu, Z. A. and Gao, B. X. and Wang, L. and Zhang, X. T. and Zhang, X. Y. and Guo, Y. F. and Chen, R. Y.},
  journal = {Phys. Rev. B},
  volume = {108},
  issue = {23},
  pages = {235109},
  numpages = {6},
  year = {2023},
  month = {Dec},
  publisher = {American Physical Society},
  doi = {10.1103/PhysRevB.108.235109},
  url = {https://link.aps.org/doi/10.1103/PhysRevB.108.235109}}

@article{iohani_2023_wsm_co3snsS2 ,
  title = {Electronic structure evolution of the magnetic Weyl semimetal ${\mathrm{Co}}_{3}{\mathrm{Sn}}_{2}{\mathrm{S}}_{2}$ with hole and electron doping},
  author = {Lohani, Himanshu and Foulquier, Paul and Le F\`evre, Patrick and Bertran, Fran{\c{c}}ois and Colson, Doroth\'ee and Forget, Anne and Brouet, V\'eronique},
  journal = {Phys. Rev. B},
  volume = {107},
  issue = {24},
  pages = {245119},
  numpages = {12},
  year = {2023},
  month = {Jun},
  publisher = {American Physical Society},
  doi = {10.1103/PhysRevB.107.245119},
  url = {https://link.aps.org/doi/10.1103/PhysRevB.107.245119}}

@article{li_2019_ti_mnbi2te4,
author = {Jiaheng Li  and Yang Li  and Shiqiao Du  and Zun Wang  and Bing-Lin Gu  and Shou-Cheng Zhang  and Ke He  and Wenhui Duan  and Yong Xu },
title = {Intrinsic magnetic topological insulators in van der Waals layered Mn$\mathrm{Bi}_2\mathrm{Te}_4$ family materials},
journal = {Science Advances},
volume = {5},
number = {6},
pages = {eaaw5685},
year = {2019},
doi = {10.1126/sciadv.aaw5685},
URL = {https://www.science.org/doi/abs/10.1126/sciadv.aaw5685},
}

@article{wang_2019_wsm_eucd2as2,
  title = {Single pair of Weyl fermions in the half-metallic semimetal $\mathrm{EuC}{\mathrm{d}}_{2}\mathrm{A}{\mathrm{s}}_{2}$},
  author = {Wang, Lin-Lin and Jo, Na Hyun and Kuthanazhi, Brinda and Wu, Yun and McQueeney, Robert J. and Kaminski, Adam and Canfield, Paul C.},
  journal = {Phys. Rev. B},
  volume = {99},
  issue = {24},
  pages = {245147},
  numpages = {9},
  year = {2019},
  month = {Jun},
  publisher = {American Physical Society},
  doi = {10.1103/PhysRevB.99.245147},
  url = {https://link.aps.org/doi/10.1103/PhysRevB.99.245147}}

@article{wang_2016_anisotropic_transport_eucd2as2,
  title = {Anisotropic transport and optical spectroscopy study on antiferromagnetic triangular lattice ${\mathrm{EuCd}}_{2}{\mathrm{As}}_{2}$: An interplay between magnetism and charge transport properties},
  author = {Wang, H. P. and Wu, D. S. and Shi, Y. G. and Wang, N. L.},
  journal = {Phys. Rev. B},
  volume = {94},
  issue = {4},
  pages = {045112},
  numpages = {6},
  year = {2016},
  month = {Jul},
  publisher = {American Physical Society},
  doi = {10.1103/PhysRevB.94.045112},
  url = {https://link.aps.org/doi/10.1103/PhysRevB.94.045112}}

@article{krishna_2018_1st_principles_of_eucd2as2,
  title = {First-principles study of electronic structure, transport, and optical properties of ${\mathrm{EuCd}}_{2}{\mathrm{As}}_{2}$},
  author = {Krishna, Jyoti and Nautiyal, T. and Maitra, T.},
  journal = {Phys. Rev. B},
  volume = {98},
  issue = {12},
  pages = {125110},
  numpages = {7},
  year = {2018},
  month = {Sep},
  publisher = {American Physical Society},
  doi = {10.1103/PhysRevB.98.125110},
  url = {https://link.aps.org/doi/10.1103/PhysRevB.98.125110}}

@article{hasan_kane_colloquium_tis,
  title = {Colloquium: Topological insulators},
  author = {Hasan, M. Z. and Kane, C. L.},
  journal = {Rev. Mod. Phys.},
  volume = {82},
  issue = {4},
  pages = {3045--3067},
  numpages = {0},
  year = {2010},
  month = {Nov},
  publisher = {American Physical Society},
  doi = {10.1103/RevModPhys.82.3045},
  url = {https://link.aps.org/doi/10.1103/RevModPhys.82.3045}}

@article{qi_zhang_tis_and_superconductors,
  title = {Topological insulators and superconductors},
  author = {Qi, Xiao-Liang and Zhang, Shou-Cheng},
  journal = {Rev. Mod. Phys.},
  volume = {83},
  issue = {4},
  pages = {1057--1110},
  numpages = {0},
  year = {2011},
  month = {Oct},
  publisher = {American Physical Society},
  doi = {10.1103/RevModPhys.83.1057},
  url = {https://link.aps.org/doi/10.1103/RevModPhys.83.1057}}

@article{
qi_zhang_ti_monopole,
author = {Xiao-Liang Qi  and Rundong Li  and Jiadong Zang  and Shou-Cheng Zhang },
title = {Inducing a Magnetic Monopole with Topological Surface States},
journal = {Science},
volume = {323},
number = {5918},
pages = {1184-1187},
year = {2009},
doi = {10.1126/science.1167747},
URL = {https://www.science.org/doi/abs/10.1126/science.1167747},
}

@article{chang_2009_tis-_optical_signature,
  title = {Optical signature of topological insulators},
  author = {Chang, Ming-Che and Yang, Min-Fong},
  journal = {Phys. Rev. B},
  volume = {80},
  issue = {11},
  pages = {113304},
  numpages = {4},
  year = {2009},
  month = {Sep},
  publisher = {American Physical Society},
  doi = {10.1103/PhysRevB.80.113304},
  url = {https://link.aps.org/doi/10.1103/PhysRevB.80.113304}
}

@article{tse_2010_kerr_faraday_effect_tis,
  title = {Giant Magneto-Optical Kerr Effect and Universal Faraday Effect in Thin-Film Topological Insulators},
  author = {Tse, Wang-Kong and MacDonald, A. H.},
  journal = {Phys. Rev. Lett.},
  volume = {105},
  issue = {5},
  pages = {057401},
  numpages = {4},
  year = {2010},
  month = {Jul},
  publisher = {American Physical Society},
  doi = {10.1103/PhysRevLett.105.057401},
  url = {https://link.aps.org/doi/10.1103/PhysRevLett.105.057401}
}

@article{grushin_2011_casimir_tis,
  title = {Tunable Casimir Repulsion with Three-Dimensional Topological Insulators},
  author = {Grushin, Adolfo G. and Cortijo, Alberto},
  journal = {Phys. Rev. Lett.},
  volume = {106},
  issue = {2},
  pages = {020403},
  numpages = {4},
  year = {2011},
  month = {Jan},
  publisher = {American Physical Society},
  doi = {10.1103/PhysRevLett.106.020403},
  url = {https://link.aps.org/doi/10.1103/PhysRevLett.106.020403}
}

@article{lopez_2011_casimir_tis,
  title = {Casimir repulsion between topological insulators in the diluted regime},
  author = {Rodriguez-Lopez, Pablo},
  journal = {Phys. Rev. B},
  volume = {84},
  issue = {16},
  pages = {165409},
  numpages = {7},
  year = {2011},
  month = {Oct},
  publisher = {American Physical Society},
  doi = {10.1103/PhysRevB.84.165409},
  url = {https://link.aps.org/doi/10.1103/PhysRevB.84.165409}
}

@article{crosse_2015_electromagnetic_greens_function_layered_tis,
  title = {Electromagnetic Green's function for layered topological insulators},
  author = {Crosse, J. A. and Fuchs, Sebastian and Buhmann, Stefan Yoshi},
  journal = {Phys. Rev. A},
  volume = {92},
  issue = {6},
  pages = {063831},
  numpages = {14},
  year = {2015},
  month = {Dec},
  publisher = {American Physical Society},
  doi = {10.1103/PhysRevA.92.063831},
  url = {https://link.aps.org/doi/10.1103/PhysRevA.92.063831}
}

@article{franca_2022_modification_of_transition_radiation_3dtis,
  title = {Modification of transition radiation by three-dimensional topological insulators},
  author = {Franca, O. J. and Buhmann, Stefan Yoshi},
  journal = {Phys. Rev. B},
  volume = {105},
  issue = {15},
  pages = {155120},
  numpages = {19},
  year = {2022},
  month = {Apr},
  publisher = {American Physical Society},
  doi = {10.1103/PhysRevB.105.155120},
  url = {https://link.aps.org/doi/10.1103/PhysRevB.105.155120}
}

@Article{zhang_2009_tis_in_bi2se3_bite3_sb2te3,
author={Zhang, Haijun
and Liu, Chao-Xing
and Qi, Xiao-Liang
and Dai, Xi
and Fang, Zhong
and Zhang, Shou-Cheng},
title={Topological insulators in Bi2Se3, Bi2Te3 and Sb2Te3 with a single Dirac cone on the surface},
journal={Nature Physics},
year={2009},
month={Jun},
day={01},
volume={5},
number={6},
pages={438-442},
issn={1745-2481},
doi={10.1038/nphys1270},
url={https://doi.org/10.1038/nphys1270}
}

@article{wu_2016_bi2se3_kerr_faraday,
author = {Liang Wu  and M. Salehi  and N. Koirala  and J. Moon  and S. Oh  and N. P. Armitage },
title = {Quantized Faraday and Kerr rotation and axion electrodynamics of a 3D topological insulator},
journal = {Science},
volume = {354},
number = {6316},
pages = {1124-1127},
year = {2016},
doi = {10.1126/science.aaf5541},
URL = {https://www.science.org/doi/abs/10.1126/science.aaf5541},
}

@Article{dziom_2017_strained_hgte_magnetoelectric,
author={Dziom, V.
and Shuvaev, A.
and Pimenov, A.
and Astakhov, G. V.
and Ames, C.
and Bendias, K.
and B{\"o}ttcher, J.
and Tkachov, G.
and Hankiewicz, E. M.
and Br{\"u}ne, C.
and Buhmann, H.
and Molenkamp, L. W.},
title={Observation of the universal magnetoelectric effect in a 3D topological insulator},
journal={Nature Communications},
year={2017},
month={May},
day={15},
volume={8},
number={1},
pages={15197},
issn={2041-1723},
doi={10.1038/ncomms15197},
url={https://doi.org/10.1038/ncomms15197}
}

@article{feynman_paradox_ti,
  title = {Feynman paradox in a spherical axion insulator},
  author = {Chyzhykova, Anastasiia and van den Brink, Jeroen and Nogueira, Flavio S.},
  journal = {Phys. Rev. B},
  volume = {113},
  issue = {11},
  pages = {115414},
  numpages = {9},
  year = {2026},
  month = {Mar},
  publisher = {American Physical Society},
  doi = {10.1103/1zq6-w193},
  url = {https://link.aps.org/doi/10.1103/1zq6-w193}}

@article{pugh_1967_static_poynting,
    author = {Pugh, Emerson M. and Pugh, George E.},
    title = {Physical Significance of the Poynting Vector in Static Fields},
    journal = {American Journal of Physics},
    volume = {35},
    number = {2},
    pages = {153-156},
    year = {1967},
    month = {02},
    issn = {0002-9505},
    doi = {10.1119/1.1973915},
    url = {https://doi.org/10.1119/1.1973915},
    }

@article{corinaldesi_1980_static_angular_momentum,
    author = {Corinaldesi, E.},
    title = {Angular momentum of a static electromagnetic field},
    journal = {American Journal of Physics},
    volume = {48},
    number = {1},
    pages = {83-83},
    year = {1980},
    month = {01},
    issn = {0002-9505},
    doi = {10.1119/1.12271},
    url = {https://doi.org/10.1119/1.12271},
    }

@article{Aguirregabiria_1981,
doi = {10.1088/0143-0807/2/3/009},
url = {https://doi.org/10.1088/0143-0807/2/3/009},
year = {1981},
month = {jul},
publisher = {},
volume = {2},
number = {3},
pages = {168},
author = {J M Aguirregabiria and A Hernandez},
title = {The Feynman paradox revisited},
journal = {European Journal of Physics},}

@ARTICLE{lombardi_feynman_disk_1983,
       author = {{Lombardi}, Gabriel G.},
        title = "{Feynman's disk paradox}",
      journal = {American Journal of Physics},
         year = 1983,
        month = mar,
       volume = {51},
       number = {3},
        pages = {213-214},
          doi = {10.1119/1.13272},
       adsurl = {https://ui.adsabs.harvard.edu/abs/1983AmJPh..51..213L}}

@article{bahder_1985_feynman_paradox,
    author = {Bahder, T. and Sak, J.},
    title = {Elementary solution to Feynman’s disk paradox},
    journal = {American Journal of Physics},
    volume = {53},
    number = {5},
    pages = {495-497},
    year = {1985},
    month = {05},
    issn = {0002-9505},
    doi = {10.1119/1.14387},
    url = {https://doi.org/10.1119/1.14387},
    }

@article{Pantazis_2017_feynman_disk,
doi = {10.1088/0143-0807/38/1/015204},
url = {https://doi.org/10.1088/0143-0807/38/1/015204},
year = {2017},
month = {nov},
publisher = {IOP Publishing},
volume = {38},
number = {1},
pages = {015204},
author = {Pantazis, George and Perivolaropoulos, Leandros},
title = {A general realistic treatment of the disk paradox},
journal = {European Journal of Physics},}

@article{Jimnez2022TheFP,
  title={The Feynman paradox and hidden momentum},
  author={J. L. Jim{\'e}nez and Ignacio Campos and J A E Roa-Neri},
  journal={European Journal of Physics},
  year={2022},
  pages={055202},
  volume={43},
  url={https://api.semanticscholar.org/CorpusID:249697799}}

@ARTICLE{lindell_biisotropic_sphere,
  author={Lindell, I.V.},
  journal={IEEE Transactions on Antennas and Propagation}, 
  title={Quasi-static image theory for the bi-isotropic sphere}, 
  year={1992},
  volume={40},
  number={2},
  pages={228-233},
  doi={10.1109/8.127408}}

@ARTICLE{brevik_1979_abraham_minkowski,
       author = {Brevik, I.},
        title = "{Experiments in phenomenological electrodynamics and the electromagnetic energy-momentum tensor}",
      journal = {Physics Reports},
         year = 1979,
        month = may,
       volume = {52},
       number = {3},
        pages = {133-201},
          doi = {10.1016/0370-1573(79)90074-7},
       adsurl = {https://ui.adsabs.harvard.edu/abs/1979PhR....52..133B}}

@article{barnett_2010_enigma_optical_momentum,
    author = {Barnett, Stephen M. and Loudon, Rodney},
    title = {The enigma of optical momentum in a medium},
    journal = {Philosophical Transactions of the Royal Society A: Mathematical, Physical and Engineering Sciences},
    volume = {368},
    number = {1914},
    pages = {927-939},
    year = {2010},
    month = {03},
    issn = {1364-503X},
    doi = {10.1098/rsta.2009.0207},
    url = {https://doi.org/10.1098/rsta.2009.0207},}

@article{voigtlander_2018_stm_review,
    author = {Voigtländer, Bert and Cherepanov, Vasily and Korte, Stefan and Leis, Arthur and Cuma, David and Just, Sven and Lüpke, Felix},
    title = {Invited Review Article: Multi-tip scanning tunneling microscopy: Experimental techniques and data analysis},
    journal = {Review of Scientific Instruments},
    volume = {89},
    number = {10},
    pages = {101101},
    year = {2018},
    month = {10},
    issn = {0034-6748},
    doi = {10.1063/1.5042346},
    url = {https://doi.org/10.1063/1.5042346},}

@article{fleming_2013_afm_review,
title = {A review of nanometer resolution position sensors: Operation and performance},
journal = {Sensors and Actuators A: Physical},
volume = {190},
pages = {106-126},
year = {2013},
issn = {0924-4247},
doi = {https://doi.org/10.1016/j.sna.2012.10.016},
url = {https://www.sciencedirect.com/science/article/pii/S0924424712006267},
author = {Andrew J. Fleming},}

@article{chighizola_2020_afm_colloidal_probes,
author = {Chighizola, Matteo and Puricelli, Luca and Bellon, Ludovic and Podestà, Alessandro},
title = {Large colloidal probes for atomic force microscopy: Fabrication and calibration issues},
journal = {Journal of Molecular Recognition},
volume = {34},
number = {1},
pages = {e2879},
doi = {https://doi.org/10.1002/jmr.2879},
url = {https://onlinelibrary.wiley.com/doi/abs/10.1002/jmr.2879},
year = {2021},
}

@article{hoang_2016_torsional_optomechanics,
  title = {Torsional Optomechanics of a Levitated Nonspherical Nanoparticle},
  author = {Hoang, Thai M. and Ma, Yue and Ahn, Jonghoon and Bang, Jaehoon and Robicheaux, F. and Yin, Zhang-Qi and Li, Tongcang},
  journal = {Phys. Rev. Lett.},
  volume = {117},
  issue = {12},
  pages = {123604},
  numpages = {5},
  year = {2016},
  month = {Sep},
  publisher = {American Physical Society},
  doi = {10.1103/PhysRevLett.117.123604},
  url = {https://link.aps.org/doi/10.1103/PhysRevLett.117.123604}
}

@article{reimann_2018_ghz_rotation,
  title = {GHz Rotation of an Optically Trapped Nanoparticle in Vacuum},
  author = {Reimann, Ren\'e and Doderer, Michael and Hebestreit, Erik and Diehl, Rozenn and Frimmer, Martin and Windey, Dominik and Tebbenjohanns, Felix and Novotny, Lukas},
  journal = {Phys. Rev. Lett.},
  volume = {121},
  issue = {3},
  pages = {033602},
  numpages = {5},
  year = {2018},
  month = {Jul},
  publisher = {American Physical Society},
  doi = {10.1103/PhysRevLett.121.033602},
  url = {https://link.aps.org/doi/10.1103/PhysRevLett.121.033602}
}

@Article{ahn_2020_ultrasensitive_torque,
author={Ahn, Jonghoon
and Xu, Zhujing
and Bang, Jaehoon
and Ju, Peng
and Gao, Xingyu
and Li, Tongcang},
title={Ultrasensitive torque detection with an optically levitated nanorotor},
journal={Nature Nanotechnology},
year={2020},
month={Feb},
day={01},
volume={15},
number={2},
pages={89-93},
issn={1748-3395},
doi={10.1038/s41565-019-0605-9},
url={https://doi.org/10.1038/s41565-019-0605-9}
}

@book{maggiore_qft,
    author = {Maggiore, Michele},
    title = {A Modern Introduction to Quantum Field Theory},
    publisher = {Oxford University Press},
    address = {Oxford},
    year = {2004},
    month = {11},
    isbn = {9780198520733},
    doi = {10.1093/oso/9780198520733.001.0001},
    url = {https://doi.org/10.1093/oso/9780198520733.001.0001},}

@book{feynman_lectures_vol2,
  author    = {Richard P. Feynman and Robert B. Leighton and Matthew Sands},
  title     = {The Feynman Lectures on Physics Vol. II},
  volume    = {},
  edition   = {},
  year      = {2011},
  publisher = {Basic Books, New Millennium Edition},
  address   = {New York}}

@book{jackson_classical_electrodynamics,
  address = {New York},
  author = {Jackson, John David},
  edition = {3rd},
  isbn = {9780471309321},
  lccn = {538.3537.8},
  publisher = {Wiley},
  title = {Classical electrodynamics},
  url = {http://cdsweb.cern.ch/record/490457},
  year = 1999}

@book{goldstein_mechanics,
author = {Goldstein, Herbert and Poole, Charles and Safko, John},
year = {2001},
month = {06},
pages = {},
edition={3rd},
publisher={Addison-Wesley},
title = {Classical Mechanics},
address={},
isbn = {0201657023}}

@book{gradshteyn2007,
  author = {Gradshteyn, I. S. and Ryzhik, I. M.},
  edition = {7th },
  isbn = {978-0-12-373637-6; 0-12-373637-4},
  mrclass = {00A22 (33-00 65-00 65A05)},
  mrnumber = {2360010 (2008g:00005)},
  publisher = {Elsevier/Academic Press},
  address={Amsterdam},
  title = {Table of integrals, series, and products},
  year = 2007}

\end{document}